# Thermoelectric and stress distributions around a smooth cavity in thermoelectric material

Zhaohang Lee[1], Yu Tang[1], Wennan Zou[1*]

(1. Institute of Engineering Mechanics / Institute for Advanced Study, Nanchang University, Nanchang 330031, China)

**ABSTRACT:** Thermoelectric materials have attracted more and more attention since they are friendly to the environment and have potentials for sustainable and renewable energy applications. As typically brittle semiconductors with low mechanical strength and always subjected to defects and damages, to clarify the stress concentration is very important in the design and implement of thermoelectric devices. The two-dimensional thermoelectric coupling problem due to a cavity embedded in an infinite isotropic homogeneous thermoelectric material, subjected to uniform electric current density or uniform energy flux, is studied, where the shape of the cavity is characterized by the Laurent polynomial, and the electric insulated and adiabatic boundary around the cavity are considered. The explicit analytic solutions of Kolosov-Muskhelishvili (K-M) potentials and rigid-body translation are carried out through a novel tactic. Comparing with the reported results, the new obtained are completely exact and possess a finite form. Some results of three typical cavities are presented to analyze the electric current densities (energy fluxes) and stresses around the tips. The main conclusions include: the distribution of thermoelectric field and stress at the tip obviously depends on the curvature of the contour and loading directions; for triangle and square with symmetrical tips, the maximum thermoelectric and stress concentration reach the maximum or minimum when the loading direction is parallel to or perpendicular to the symmetry axis of the tip, which is distinct to the extremum characteristics of pentagram with bimodal of curvature around the tip; the maximum thermoelectric and stress concentration appear near the maximum curvature point for most load directions, but not at the maximum curvature point.

**Key words** Thermoelectric material, K-M potentials, Laurent polynomial, arbitrarily-shaped cavity

## 1 Introduction

Thermoelectric materials (TEMs) are widely used in industry and engineering as special functional materials to realize the conversion between heat energy and electric energy, such as coolers, power generations, thermal-energy sensors and spaceships (O'Brien *et al*. 2008; Hsu *et al*. 2011; Kopparthy *et al*. 2012; Wang *et al*. 2013). TEMs are eco-friendly materials; in comparing with traditional cooling and heating systems, thermoelectric equipment uses no refrigerants or working fluids, and the emissions of greenhouse gases during its use are negligible. However, the cavities caused by design or manufacturing are unavoidable in thermoelectric materials, the effects of resulting discontinuity must be paid attention on the properties of thermoelectric materials.

Cavities are the major charge carriers for the *p*-type thermoelectric materials (Yao *et al*. 2018), the perturbation effect caused by the cavities inevitably affects the heat and electrical conduction behavior. Pang *et al*. (2018) found that the electric and energy fields intensity factors around the crack tip were closely related to the shape characteristics of the cavity and the crack length. Song *et al*. (2019b) pointed out that the electrical and thermal conductivity around the cavity depends largely on the size and shape of the cavity. Song *et al*. (2021) showed that the cavity has the same effect on effective electric and thermal conductivities. In addition, the presence of nano-cavities can improve thermoelectric performance by increasing phonon scattering (Martinez *et al*. 2011). Galli *et al*. (2010) etched nanopores onto the silicon film, which greatly reduces the thermal conductivity of the film and has only a small effect on its good electrical properties, thus improves the thermoelectric conversion efficiency. Yang *et al*. (2015) found that the high phonon scattering caused by the high density grain boundary can enhance the thermoelectric properties. Hua *et al*. (2016) presented that the smaller the radius of the cavity, the smaller the effective thermal conductivity for a given porosity.

* Corresponding author: Wennan Zou, email: zouwn@ncu.edu.cn; Zhaohang Lee, email: leezh@email.ncu.edu.cn.

On the other hand, for the surrounding matrix as a brittle material, the change of the mechanical behavior caused by the cavity can not be ignored, and the complex potential theory has been proved to be a powerful formulation in the context of 2D isotropic conduction and elasticity (Muskhelishvili, 1953; England, 1971; Lu, 1995; Sadd, 2005). Song, Gao and Li (2015) studied the two-dimensional elliptical crack problem in thermoelectric materials, the fields of heat flow, electric current, and stress at the crack tip under remote electric current and heat flow are given in the form of complex variables. Zhang and Wang (2016) obtained explicit solutions for an elliptical cavity subjected to uniform electric current density and energy flux at infinity, they found that the stress concentration at the cavity rim is closely related to the ratio of the major axis to the minor one of ellipse. Wang and Wang (2017) discussed the effect of biaxial loading on the inclined elliptical hole, and the results show that the maximum thermoelectric concentration is obtained when the major axis is perpendicular to the loading direction, and the stress concentration reaches the maximum when the major axis is parallel to the loading direction. Zhang *et al*. (2017) studied the effects of elliptic geometry and heat conductivity on thermoelectric and stress fields for cavity problem. Yu *et al*. (2018) analyzed the stress intensity factors of arc-shaped crack and found that it is related to the direction of loading, material properties and crack shape. Yu *et al*. (2019) presented the closed-form solutions for arbitrarily-shaped cavity defined by a polynomial conformal mapping. Song *et al*. (2019a) considered the contribution of surface elasticity to the stress distribution around the cavity. It is found that the studies of cavity in thermoelectric materials mainly focus on elliptical shape, and the general solution of complex cavities whose mapping polynomials have more than three terms is lacking.

In this paper, the plane problem of a non-elliptical cavity embedded in thermoelectric materials solely applied by uniform electric current density or uniform energy flux at infinity is investigated, under consideration of the electrical and thermal insulation boundary; and the explicit solutions are obtained by the tactics proposed in our previous work (Zou and He, 2018). The rest of this paper is organized as follows. In Section 2, the basic theory of the problem is formulated. The electric, temperature and stress field are briefly presented by complex theory, where the K–M potentials are divided in two parts: one are the basic functions to express the thermal dislocation and the relative rigid-body translation of the cavity to the matrix, another are the perturbance functions to satisfy the traction-free condition on the boundary of the cavity and to ensure the zero value at infinity of the matrix. In second 3, the general explicit solutions of the K–M potentials are obtained by a novel tactic shown Appendix A, and the validity of the solutions is proved by using our method to compare the previous results. In second 4, three typical shapes, triangle, square and pentagram, are considered, and thermoelectric fields and stresses along the contour of the cavity are discussed. Some concluding remarks are drawn in Section 5.

## 2 Formulation of the problem, including thermoelectric field solutions and pretreatment of K-M potentials

### 2.1 Description of the problem

In two-dimensional (2D) space, consider an infinite thermoelectric material $\Omega$ with a cavity, which according to the Riemann mapping theorem can be uniquely expressed by a Laurent series (see, e.g., Henrici, 1986; Zou *et al*. 2010)

$$z = z(w) = h + R\phi(w) = h + R\left(w + \sum_{k=1}^{\infty} b_k w^{-k}\right), |w| \geq 1, \tag{1}$$

where the complex variable $z$ in the physical plane is expressed in the Cartesian coordinates $(x_1, x_2)$ as $z = x_1 + \iota x_2$, with $\iota = \sqrt{-1}$ being the unit imaginary number. For any simply connected cavity, the above formula maps the exterior of the cavity onto the exterior of the unit circle, with $h$ as an interior point, $R$ being a positive real parameter to indicate the size, and $b_k$ the complex variable parameters to describe the shape of the cavity. For an arbitrary accuracy requirement, the Laurent polynomial with a limited number of terms $N$ (England, 1971)

$$t = h + R\phi(\eta) = h + R\left(\eta + \sum_{k=1}^{N} b_k \eta^{-k}\right), |\eta| = 1 \tag{2}$$

can be used to map the point $\eta$ on the unit circle of the image plane to the point $t$ on the boundary of the cavity, where $h = 0$ is taken in this paper.

Now we assume that the matrix containing a cavity with a traction-free, electrically and thermally insulated boundary is subjected to uniform current density or uniform energy flux at infinity. The action is described by

$$\boldsymbol{\sigma}(z) = \mathbf{0};\ \boldsymbol{J}_e(z) = J_e^{\infty}\boldsymbol{n}_{\beta_e};\ \boldsymbol{J}_u(z) = J_u^{\infty}\boldsymbol{n}_{\beta_u}, \qquad z \to \infty \tag{3}$$

$$\boldsymbol{\sigma}_n(t) = \boldsymbol{\sigma}(t)\cdot\boldsymbol{n} = \mathbf{0}; J_n^e(t) \equiv \boldsymbol{J}_e(t)\cdot\boldsymbol{n} = 0;\ J_n^u(t) \equiv \boldsymbol{J}_u(t)\cdot\boldsymbol{n} = 0, \qquad t \in \Gamma. \tag{4}$$

where $\boldsymbol{\sigma}$, $\boldsymbol{J}_e$ and $\boldsymbol{J}_u$ indicate stress, current density vector and energy flux vector, respectively, $\boldsymbol{n}_{\beta_e}$ and $\boldsymbol{n}_{\beta_u}$, also denoted by complex variables $n_{\beta_e} = e^{\iota\beta_e}$ and $n_{\beta_u} = e^{\iota\beta_u}$, are the unit vectors indicating the directions of the current density and energy flux, with $\beta_e$ and $\beta_u$ being the included angles between the current density and energy flux directions and the $x_1$-axis, respectively, as shown in Fig. 1; $\boldsymbol{n}$ is the normal of the boundary $\Gamma$ of the matrix to the cavity.

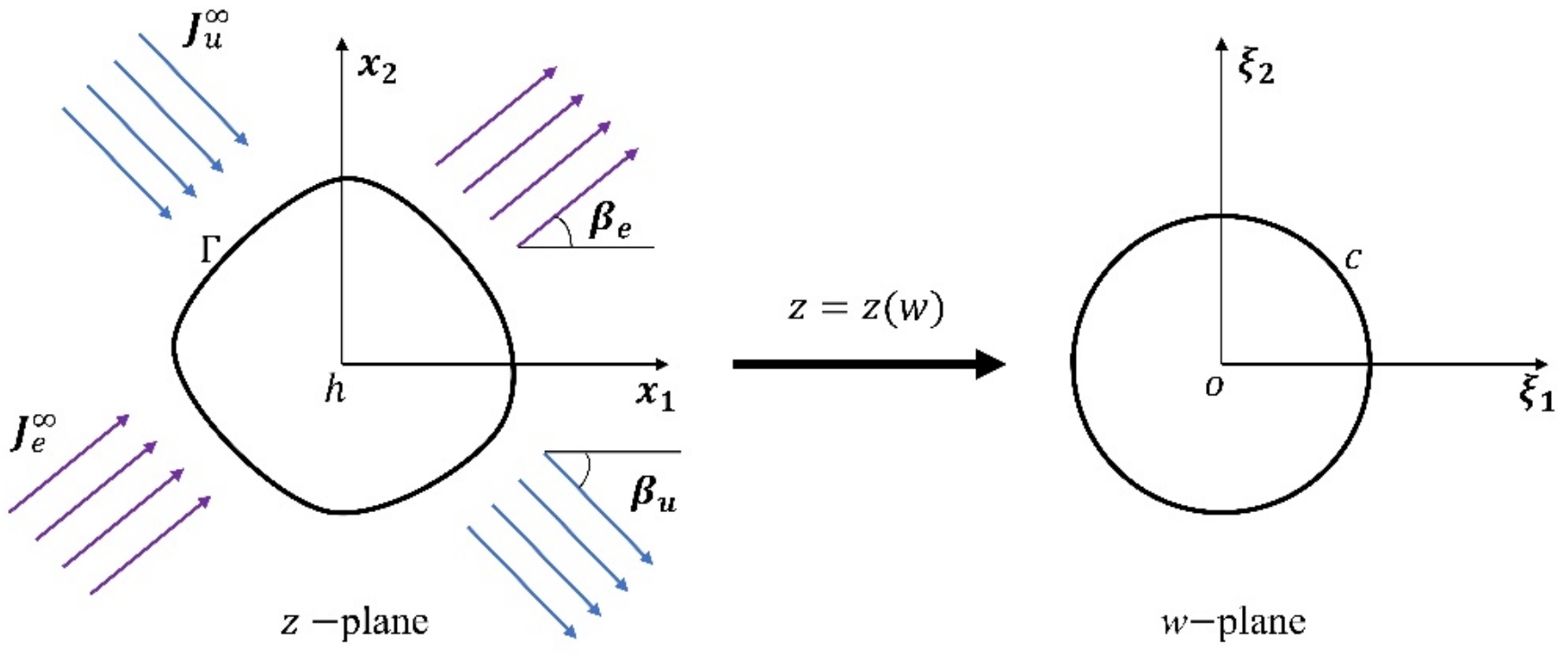

**Figure 1.** Conformal mapping of the cavity.

### 2.2 Thermoelectric fields and their solutions

#### 2.2.1 Formulation of thermoelectric problem

For a homogeneous and isotropic thermoelectric substance, the constitutive equations can be given by (Perez-Aparicio *et al.* 2007)

$$\boldsymbol{J}_e = -\sigma\nabla V - \sigma\varepsilon\nabla T;\ \boldsymbol{J}_q = -k\nabla T + \varepsilon T\boldsymbol{J}_e, \tag{5}$$

where $\boldsymbol{J}_e$ and $\boldsymbol{J}_q$ are current density vector and heat flux vector, $V$ and $T$ represent electric field and temperature field in the material with properties of electrical conductivity $\sigma$, thermal conductivity $k$ and Seebeck coefficient $\varepsilon$. The first formula superimposes the Seebeck effect on Ohm's law, while the second one consists of the Thompson and Peltier effects. These three coupled effects are called the thermoelectric effects. In addition, the equilibrium equations in the stationary state and without free electric charge and heat source take the form (Perez-Aparicio *et al.* 2007; Lee, 2016)

$$\nabla\cdot\boldsymbol{J}_e = 0;\ \nabla\cdot\boldsymbol{J}_q + \boldsymbol{J}_e\cdot\nabla V = 0. \tag{6}$$

Introducing the energy flux $\boldsymbol{J}_u = \boldsymbol{J}_q + V\boldsymbol{J}_e$ (Yang *et al.* 2013), which means that the transmission of heat and electricity in materials are expressed in terms of energy, then $(6)_2$ can simplify to

$$\nabla \cdot \boldsymbol{J}_u = 0. \tag{7}$$

Further introduce a new variable $H = V + \varepsilon T$ (Zhang and Wang, 2013), which represents the total electric potential and is independent of the whole thermoelectric problem, we have the decoupling thermoelectric equations (Zhang and Wang 2013; Yu *et al.*, 2019) of $T$ and $H$ instead of those of $T$ and $V$, that is,

$$\boldsymbol{J}_e = -\sigma \nabla H; \ \boldsymbol{J}_u = H\boldsymbol{J}_e - k\nabla T \tag{8}$$

and

$$\nabla^2 H = 0; \ k\nabla^2 T + \sigma(\nabla H)^2 = 0. \tag{9}$$

It's obvious that $H$ is a harmonic function satisfying Laplace's equation, such that the key to solve (9) lies in the treatment of the nonlinear term in $(9)_2$.

Back to the 2D situation we care about, the thermoelectric problem can be solved effectively by complex variable method, the general solution of $(9)_1$ is (Song *et al.* 2015)

$$H = \mathrm{Re}[f(z)] \equiv \frac{1}{2}\left[f(z) + \overline{f(z)}\right] \tag{10}$$

with $f(z)$ being an undetermined analytic function, $\mathrm{Re}[\cdot]$ indicates the real part of a complex variable $(\cdot)$, "$\overline{(\cdot)}$" denotes the conjugation of a complex variable $(\cdot)$. Combining (10) with $(9)_2$ yields

$$\nabla^2 T = -\frac{\sigma}{k}(\nabla H)^2 = -\frac{\sigma}{k} f'(z)\overline{f'(z)}. \tag{11}$$

Its solution can be constructed by a particular part $T_p$ and a general part $T_g$ in the form (Zhang and Wang 2016)

$$T = T_p + T_g = -\frac{\sigma}{4k} f(z)\overline{f(z)} + \mathrm{Re}g(z), \tag{12}$$

where $g(z)$ is another undetermined analytic function. Substitution of (10) into $(8)_1$ yields the complex current density

$$J_1^e - \iota J_2^e = -\sigma\left(\frac{\partial H}{\partial x_1} - \iota\frac{\partial H}{\partial x_2}\right) = -\sigma\frac{\partial H}{\partial z} = -\sigma f'(z). \tag{13}$$

From equations $(8)_2$, (10) and (12), the energy flux can be expressed by an analytic function, say

$$J_1^u - \iota J_2^u = -\frac{\sigma}{2} f'(z)f(z) - kg'(z) = -\frac{d}{dz}\left[\frac{\sigma}{4}f^2(z) + kg(z)\right]. \tag{14}$$

In addition, for electrically and thermally insulated boundary with normal $n = n_1 + \iota n_2 = -\iota\frac{dt}{ds}$, where $s$ indicates the arc length coordinate along the boundary in an anti-clockwise way, integrating (13) and (14) result in the boundary relations for the analytic functions

$$\mathrm{Im}[\sigma f(t)] = -\int J_n^e(s)ds = 0, \qquad t(s) \in \Gamma, \tag{15}$$

$$\mathrm{Im}\left[\frac{\sigma}{4}f^2(t) + kg(t)\right] = -\int J_n^u(s)ds = 0, \qquad t(s) \in \Gamma, \tag{16}$$

where $\mathrm{Im}[\cdot]$ indicates the imaginary part of a complex number $[\cdot]$, $J_n^e$ and $J_n^u$ stand for the electric flux and energy flux in the normal direction of the boundary. In the following, according to the conditions at infinity and on the boundary of the cavity, we will determine two analytic functions $f(z)$ and $g(z)$ from equations (13)-(16), respectively.

### 2.2.2 Solutions of electric and temperature fields for the cavity problem

The complex function $f(z)$ can be broken down into two parts (Zhang and Wang, 2016; Yu *et al.*, 2019)

$$f(z) = f_r(z) + f_c(z) \tag{17}$$

where the basic part $f_r(z)$ is given by the remote current density $J_e^\infty$, $f_c(z)$ is the complementary part to satisfy the condition on the boundary of the cavity, and have $f_c(\infty) = f_c'(\infty) = 0$. Taking the limit $z \to \infty$, we have from $(3)_2$ and (13) that

$$f_r(z) = -\frac{J_e^\infty e^{-\iota\beta_e}}{\sigma} z. \tag{18}$$

Substituting $(4)_2$, (17) and (18) into (15) yields

$$f_c(t) - \overline{f_c(t)} = \frac{J_e^\infty}{\sigma}\left[e^{-\iota\beta_e}t - e^{\iota\beta_e}\bar{t}\right], \qquad t \in \Gamma. \tag{19}$$

For the cavity characterized by (2), and considering the holomorphic property of $f_c(t)$ at infinity, we get the solution in terms of the complex variable $w$ in the image plane as

$$f_c(z(w)) = \frac{J_e^\infty R}{\sigma}\left(e^{-\iota\beta_e}\sum_{k=1}^{N} b_k w^{-k} - e^{\iota\beta_e}w^{-1}\right), |w| \geq 1. \tag{20}$$

In the following, when not causing confusion, we abbreviate $f(z(w))$ to $f(w)$, $g(z(w))$ to $g(w)$, etc. Combining (18) and (20) into (17) yields

$$f(w) = -\frac{J_e^\infty R}{\sigma}\left(e^{-\iota\beta_e}w + e^{\iota\beta_e}w^{-1}\right). \tag{21}$$

Similarly, $g(z)$ is also divided into two parts (Zhang and Wang, 2016; Yu *et al*., 2019)

$$g(z) = g_r(z) + g_c(z), \tag{22}$$

it can be seen from (14) that the basic part takes the form $g_r(z) = C_1 z^2 + C_2 z$, where $C_1$ and $C_2$ are complex coefficients to be determined by $J_e^\infty/\beta_e$ and $J_u^\infty/\beta_u$. Combining it with (3)₃ yields $C_1 = -\frac{J_e^{\infty 2} e^{-2\iota\beta_e}}{4k\sigma}$ and $C_2 = -\frac{J_u^\infty e^{-\iota\beta_u}}{k}$ and so

$$g_r(z) = -\frac{J_e^{\infty 2} e^{-2\iota\beta_e}}{4k\sigma}z^2 - \frac{J_u^\infty e^{-\iota\beta_u}}{k}z. \tag{23}$$

By substituting of (4)₃, (22) and (23) into boundary constraint (16), the complementary part $g_c(z)$ should satisfy

$$g_c(t) - \overline{g_c(t)} = \frac{J_e^{\infty 2}}{4k\sigma}\left(e^{-2\iota\beta_e}t^2 - e^{2\iota\beta_e}\bar{t}^2\right) + \frac{J_u^\infty}{k}\left(e^{-\iota\beta_u}t - e^{\iota\beta_u}\bar{t}\right), \qquad t \in \Gamma; \tag{24}$$

On the boundary (2) of the cavity, and considering the holomorphic property of $g_c(t)$ at infinity in the image plane, we can get the solution with respect to the complex variable $w$ as

$$g_c(w) = \frac{J_e^{\infty 2} R^2}{4k\sigma}\left[2\sum_{k=1}^{N} e^{-2\iota\beta_e} b_k w^{-k+1} + e^{-2\iota\beta_e}\left(\sum_{k=1}^{N} b_k w^{-k}\right)^2 - e^{2\iota\beta_e}w^{-2} + 2e^{-2\iota\beta_e}b_1\right]$$
$$+\frac{J_u^\infty R}{k}\left[e^{-\iota\beta_u}\sum_{k=1}^{N} b_k w^{-k} - e^{\iota\beta_u}w^{-1}\right], |w| \geq 1. \tag{25}$$

Substituting of (23) and (25) into (22) yields

$$g(w) = -\frac{J_e^{\infty 2} R^2}{4k\sigma}\left(e^{-2\iota\beta_e}w^2 + e^{2\iota\beta_e}w^{-2}\right) - \frac{J_u^\infty R}{k}\left(e^{-\iota\beta_u}w + e^{\iota\beta_u}w^{-1}\right). \tag{26}$$

The complex analytic functions $f(w)$ and $g(w)$ have compact form in the image plane and are independent of the shape of the cavity, which coincide with (22) and (23) of Song *et al.* (2019a). Furthermore, electric field and temperature field caused by current density and/or energy flux at infinity can be obtained from (13) and (14) when $f(w)$ and $g(w)$ are known.

### 2.3 Formulation of elastic fields for the cavity problem of thermoelectric material

The remaining problem is to determine the elastic fields due to the non-uniform temperature. Combining the constitutive equations, equilibrium equations and compatibility equations, the Airy stress function $\Phi$ requires that (Heinz, 1976, page 29; Song *et al.* 2015)

$$\nabla^4\Phi + \lambda\nabla^2 T = 0, \tag{27}$$

where $\lambda$ is a material constant determined by the elastic modulus $E$, linear expansion coefficient $\alpha$ and Poisson's ratio $\nu$ as

$$\lambda = \begin{cases} E\alpha, & \text{plane stress;} \\ \dfrac{E\alpha}{1-\nu}, & \text{plane strain.} \end{cases} \tag{28}$$

Similar to the construction of the temperature, making use of the solution (12), the solution of $\Phi$ can be divided into a particular solution $\Phi_p$ and a general solution $\Phi_g$ (Muskhelishvili, 1953; Zhang and Wang, 2016), say

$$\Phi = \Phi_p + \Phi_g = \frac{\lambda\sigma}{16k}F(z)\overline{F(z)} + \frac{1}{2}\left[\bar{z}\varphi(z) + z\overline{\varphi(z)} + \theta(z) + \overline{\theta(z)}\right], \tag{29}$$

where $\varphi(z)$ and $\theta(z)$ are two complex functions to be solved, and

$$F(z)=\int_{z_0}^{z} f(y)dy=-\frac{J_e^{\infty}R^2}{\sigma}\left(e^{\iota\beta_e}-b_1e^{-\iota\beta_e}\right)\ln w+P_F(w), \tag{30}$$

where $P_F(w)$ is a polynomial with respect to complex variable $w$ in the image plane, and $z_0$ is usually taken to be a point on the boundary of the cavity.

The stress and the displacement relative to point $z_0$ can be expressed as (Heinz, 1976)

$$\begin{cases}\sigma_{11}+\sigma_{22}=4\dfrac{\partial^2\Phi}{\partial z\partial\bar{z}}=2\left[\varphi'(z)+\overline{\varphi'(z)}\right]+\dfrac{\lambda\sigma}{4k}f(z)\overline{f(z)},\\ \sigma_{22}-\sigma_{11}+2\iota\tau_{12}=4\dfrac{\partial^2\Phi}{\partial z^2}=2[\bar{z}\varphi''(z)+\psi'(z)]+\dfrac{\lambda\sigma}{4k}f'(z)\overline{F(z)},\end{cases} \tag{31}$$

$$U\equiv u_1+\iota u_2=\frac{1}{2\mu}\left[\kappa\varphi(z)-z\overline{\varphi'(z)}-\overline{\psi(z)}-\frac{\lambda\sigma}{8k}\overline{f(z)}F(z)\right]+\alpha'\int_{z_0}^{z}\mathrm{Re}[g(y)]dy, \tag{32}$$

where $\psi(z)=\theta'(z)$, and $\mu$ is the shear modulus of the material, $\kappa$ and $\alpha'$ are parameters associated with Poisson's ratio $\nu$ and linear expansion coefficient $\alpha$, respectively

$$\kappa=\begin{cases}\dfrac{3-\nu}{1+\nu}, & \text{plane stress},\\ 3-4\nu, & \text{plane strain};\end{cases}\quad \alpha'=\begin{cases}\alpha, & \text{plane stress},\\ (1+\nu)\alpha, & \text{plane strain}.\end{cases} \tag{33}$$

The traction $f^t$ on the boundary with arc-length coordinate $s$ can be indicated by

$$f^t=f_1^t+\iota f_2^t=-\iota\frac{d}{ds}\left[\varphi(t)+t\overline{\varphi'(t)}+\overline{\psi(t)}+\frac{\lambda\sigma}{8k}\overline{f(t)}F(t)\right], t(s)\in\Gamma. \tag{34}$$

Following Florence and Goodier (1960) and Zou and He (2018), the K-M potentials consist of the basic potentials $\varphi_0(z)$ and $\psi_0(z)$ and the perturbation potentials $\varphi_p(z)$ and $\psi_p(z)$

$$\varphi(z)=\varphi_0(z)+\varphi_p(z),\qquad \psi(z)=\psi_0(z)+\psi_p(z). \tag{35}$$

The basic potentials due to the thermal dislocation of temperature inhomogeneity and the rigid-body translation of the cavity relative to the matrix could to be represented in the form (Zhang and Wang, 2016)

$$\varphi_0(w)=RA\ln w,\ \psi_0(w)=2\mu\bar{U}_0+RB(w)\ln w, \tag{36}$$

where $A$ is a constant, $B(w)$ is assumed to be a function due to the nonlinearity of $(9)_2$, $U_0$ denotes the rigid-body translation, which will be solved in Appendix A. The perturbation potentials caused by the boundary compatibility of the cavity take the form

$$\varphi_p(w)=R\sum_{k=1}^{\infty}\alpha_k w^{-k},\qquad \psi_p(w)=R\sum_{k=1}^{\infty}\beta_k w^{-k},\qquad |w|\geq 1. \tag{37}$$

where $\alpha_k$ and $\beta_k$ are two unknown coefficients to be solved.

First, the single-valued conditions of displacement (32) and resultant force (34) along the boundary of the cavity (Zhang and Wang, 2016; Zhang *et al.*, 2017; Wang and Wang, 2017), namely

$$0=[U]_{\eta=e^{0\iota}}^{\eta=e^{2\pi\iota}}=\frac{1}{2\mu}\left[\kappa\varphi(t)-t\overline{\varphi'(t)}-\overline{\psi(t)}-\frac{\lambda\sigma}{8k}\overline{f(t)}F(t)\right]_{\eta=e^{0\iota}}^{\eta=e^{2\pi\iota}}+\alpha'\oint_{\Gamma}g(t)dt, \tag{38}$$

$$0=\iota\oint_{\Gamma}f^t ds=\left[\varphi(t)+t\overline{\varphi'(t)}+\overline{\psi(t)}+\frac{\lambda\sigma}{8k}\overline{f(t)}F(t)\right]_{\eta=e^{0\iota}}^{\eta=e^{2\pi\iota}} \tag{39}$$

result in

$$0=\frac{\pi\iota R}{\mu}\left[\kappa A+\overline{B(1)}-\frac{\lambda J_e^{\infty 2}R^2\cos\beta_e}{4k\sigma}\left(e^{\iota\beta_e}-b_1e^{-\iota\beta_e}\right)\right]+2\pi\iota\alpha'\left[\frac{J_e^{\infty 2}R^3}{2k\sigma}e^{-2\iota\beta_e}b_2-\frac{J_u^{\infty}R^2}{k}\left(e^{\iota\beta_u}-e^{-\iota\beta_u}b_1\right)\right], \tag{40}$$

$$0=2\pi\iota R\left[A-\overline{B(1)}+\frac{\lambda J_e^{\infty 2}R^2\cos\beta_e}{4k\sigma}\left(e^{\iota\beta_e}-b_1e^{-\iota\beta_e}\right)\right], \tag{41}$$

which yield

$$A=\sigma_{u0}\left(e^{\iota\beta_u}-e^{-\iota\beta_u}b_1\right)-\sigma_{e0}b_2e^{-2\iota\beta_e}, \tag{42}$$

$$B(1)=\bar{A}+2\sigma_{e0}\cos\beta_e\left(e^{-\iota\beta_e}-\bar{b}_1e^{\iota\beta_e}\right), \tag{43}$$

with

$$\sigma_{e0}=\frac{\mu R^2\alpha' J_e^{\infty 2}}{k(\kappa+1)\sigma}=\frac{\lambda J_e^{\infty 2}R^2}{8k\sigma},\sigma_{u0}=\frac{2\mu R\alpha' J_u^{\infty}}{k(\kappa+1)}. \tag{44}$$

In the loop calculations, all polynomial terms naturally vanish.

Further, the resultant force must vanish everywhere on the boundary of the cavity due to the traction-free condition $(4)_1$, which means

$$\varphi(t)+t\overline{\varphi'(t)}+\overline{\psi(t)}+\frac{\lambda\sigma}{8k}\overline{f(t)}F(t)=0,\qquad t\in\Gamma. \tag{45}$$

Consider the balance of all multi-valued terms in (45), that is

$$A\ln\eta-\overline{B(\eta)}\ln\eta+\sigma_{e0}\left(e^{-\iota\beta_e}\eta+e^{\iota\beta_e}\eta^{-1}\right)\left(e^{\iota\beta_e}-b_1e^{-\iota\beta_e}\right)\ln\eta=0,\qquad t\in\Gamma. \tag{46}$$

yields the solution

$$B(w)=\bar{A}+\sigma_{e0}\left(e^{-\iota\beta_e}-\bar{b}_1e^{\iota\beta_e}\right)\left(e^{-\iota\beta_e}w+e^{\iota\beta_e}w^{-1}\right). \tag{47}$$

The expressions of (42) and (47) are the same as (43) of Song *et al.* (2019a), except a typo that the $m_1$ in $(43)_2$ should be $\bar{m}_1$. We can see that that parameters $A$ and $B(w)$ are related to material properties, the shape characteristic of the cavity and loads at infinity. It is worth noting that $A$ and $B(w)$ are only related to $b_1$ and $b_2$ among the shape parameters of the cavity.

Substituting of (35), (36), (42) and (47) into (45) yields the constraint condition of the perturbance potentials on the boundary

$$\varphi_p(\eta)+\frac{t(\eta)}{\overline{t'(\eta)}}\overline{\varphi_p'(\eta)}+\overline{\psi_p(\eta)}=-2\mu U_0-\frac{\bar{A}t(\eta)}{\overline{t'(\eta)}}R\eta-R\sigma_{e0}\left(\frac{1}{2}\eta+\frac{1}{2}b_1\eta^{-1}+\sum_{k=2}^{N}\frac{2k^2}{k^2-1}b_k\eta^{-k}\right.$$
$$\left.+\frac{1}{2}e^{-2\iota\beta_e}\eta^3+e^{-2\iota\beta_e}\sum_{k=2}^{N}\frac{k}{k-1}b_k\eta^{-k+2}+e^{2\iota\beta_e}\sum_{k=1}^{N}\frac{k}{k+1}b_k\eta^{-k-2}\right),|\eta|=1,t(\eta)\in\Gamma. \tag{48}$$

The perturbation potentials $\varphi_p(w)$ and $\psi_p(w)$ above can be solved following the method proposed by Zou and He (2018), and the detailed derivation is presented in Appendix A.

## 3 Explicit analytical solutions and comparison

### 3.1 Analytical solutions of K-M potentials in a finite form

According to Appendix A, the perturbation potentials $\varphi_p(w)$ and $\psi_p(w)$ can be characterized by a negative power polynomial with finite terms when the cavity has a finite expression by (2), namely $\varphi_p(w)$ has maximal negative power $N+2$ while $\psi_p(w)\phi'(w)$ has maximal negative power $N+4$. The total K-M potentials have formulae

$$\frac{\varphi(w)}{R}=A\ln w+\sum_{k=1}^{N+2}\alpha_k w^{-k},\frac{\psi(w)}{R}=B(w)\ln w+2\mu\frac{\bar{U}_0}{R}+\frac{\sum_{k=1}^{N+4}\beta_k w^{-k}}{\phi'(\eta)}, \tag{49}$$

where $A$ and $B(w)$ are given by (42) and (47), the rigid-body translation $U_0$ and other coefficients $\{\alpha_k,\ k=1,\ldots,N+2\},\{\beta_k,k=1,\ldots,N+4\}$ are presented as follows.

Introduce the shape parameters $\{g_k,\ k=1,\ldots,N\}$

$$g_N=b_N,\qquad g_{N-1}=b_{N-1}, \tag{50.1}$$

$$g_{N-k}=b_{N-k}+\sum_{l=1}^{k-1}l g_{N-k+l+1}\bar{b}_l,\qquad k=2,\ldots,N-1; \tag{50.2}$$

and the loading parameters $\{c_k,\ k=1,\ldots,N+2\}$

$$\begin{cases} c_1=-\bar{A}g_2-\sigma_{e0}\left(\frac{3}{2}b_3e^{-2\iota\beta_e}+\frac{1}{2}b_1\right),c_2=-\bar{A}g_3-\sigma_{e0}\left(\frac{4}{3}b_4e^{-2\iota\beta_e}+\frac{8}{3}b_2\right),\\ c_k=-\bar{A}g_{k+1}-\sigma_{e0}\left(\frac{k+2}{k+1}b_{k+2}e^{-2\iota\beta_e}+\frac{2k^2}{k^2-1}b_k+\frac{k-2}{k-1}b_{k-2}e^{2\iota\beta_e}\right),k=3,\ldots,N-2,\\ c_{N-1}=-\bar{A}g_N-\sigma_{e0}\left[\frac{2(N-1)^2}{N(N-2)}b_{N-1}+e^{2\iota\beta_e}\frac{N-3}{N-2}b_{N-3}\right],c_N=-\sigma_{e0}\left(\frac{2N^2}{N^2-1}b_N+\frac{N-2}{N-1}b_{N-2}e^{2\iota\beta_e}\right),\\ c_{N+1}=-\sigma_{e0}\frac{N-1}{N}b_{N-1}e^{2\iota\beta_e},c_{N+2}=-\sigma_{e0}\frac{N}{N+1}b_Ne^{2\iota\beta_e}; \end{cases} \tag{51}$$

the coefficients $\{\alpha_k,\ k=1,\ldots,N+2\}$ of the first perturbance potential $\varphi_p(w)$ are solved from the linear

equations

$$\alpha_k = c_k + \sum_{l=1}^{N-k-1} l\bar{\alpha}_l g_{k+l+1}, \qquad k = 1, 2, \dots, N+2; \tag{52}$$

which can be solved in an iterative way as shown in (A.8) - (A.9), while $U_0$ is given by

$$2\mu\frac{U_0}{R} = \sum_{l=1}^{N-1} l g_{l+1}\bar{\alpha}_l - \bar{A}g_1 - 2b_2\sigma_{e0}e^{-2\iota\beta}. \tag{53}$$

The coefficients $\{\beta_k, k = 1, \dots, N+4\}$ of the second perturbance potential $\psi_p(w)$ are gotten from the equation

$$\begin{aligned}\sum_{k=1}^{N+4}\frac{\beta_k}{w^k} = &\sum_{k=1}^{N+2}\frac{1}{w^k}\sum_{l=k}^{N+2} l\left(b_l\bar{\alpha}_{l-k+1} + \alpha_l\bar{b}_{l-k+1}\right) + \sum_{k=1}^{N+2}\frac{k\alpha_k}{w^{k+2}} - \frac{A}{w^2} + \frac{2\mu\bar{U}_0}{R}\sum_{k=1}^{N}\frac{kb_k}{w^{k+1}}\\ &-\sigma_{e0}\left[\frac{1}{2w} - \sum_{k=1}^{N-1}\frac{1}{w^k}\sum_{p=k+1}^{N}\frac{2(p-k+1)^2}{(p-k+1)^2-1}pb_p\bar{b}_{p-k+1} - \frac{1}{2}\sum_{k=1}^{N}\frac{kb_k}{w^{k+2}} - \frac{\bar{b}_1}{2}\sum_{k=1}^{N}\frac{kb_k}{w^k}\right.\\ &+e^{2\iota\beta_e}\left(\frac{1}{2w^3} - \frac{1}{2}\sum_{k=1}^{N}\frac{kb_k}{w^{k+4}} - 2\bar{b}_2\sum_{k=1}^{N}\frac{kb_k}{w^{k+1}} - \sum_{k=1}^{N}\frac{1}{w^k}\sum_{p=1}^{N+k-3}\frac{p-k+3}{p-k+2}pb_p\bar{b}_{p-k+3}\right)\\ &\left.-\sum_{k=1}^{N-2}\frac{e^{-2\iota\beta_e}}{w^k}\sum_{p=k+2}^{N}\frac{p-k-1}{p-k}pb_p\bar{b}_{p-k-1}\right].\end{aligned} \tag{54}$$

In the thermoelectric problem of a cavity under the remote loadings of current density or heat flux, the nonlinearity of $H$ in $(9)_2$ results in a non-harmonic temperature field. Thus, besides the thermal dislocations induced by both current density or heat flux, the thermal force in equilibrium equation (27), coming from the remote current density, also introduce a multi-valuedness of the K-M potentials, consisting of the major complexity of stress field. The formulae (21), (26) and (49) constitutes the solution to this problem with the cavity characterized by a Laurent polynomial: it has finite form with the highest power slightly greater than that of the Laurent polynomial. In the reported literature (Song *et al.* 2015; Zhang and Wang 2016; Wang and Wang 2017; Zhang *et al.* 2017; Yu *et al.* 2019; Song *et al.* 2019a, b), except most studies focusing on the elliptical shape, the existing solutions of K-M potentials for non-elliptical shapes (Yu *et al.*, 2019; Song *et al.*, 2019a, b) are expressed by truncated series, and the shapes of illustrated examples are described by mapping polynomials no more than three terms. The results obtained in this paper are completely exact and operable for more complicated shapes, since we consider the rigid-body translation of the cavity relative to the matrix and make use of the novel tactics first proposed by Zou and He (2018).

For the shapes with few terms in the Laurent polynomials, the equation (52) can be solved explicitly, and all fields get their explicit expressions. In the following, the solutions of elliptical and simple non-elliptical cavities are presented and compared with the existing results.

### 3.2 Explicit solutions of an elliptical cavity and comparison

For an elliptical cavity characterized by $t(\eta) = R(\eta + b_1\eta^{-1}), |\eta| = 1$, the explicit solutions obtained by our formulas can be expressed as

$$\begin{cases}\dfrac{\varphi(w)}{R} = A\ln w - \dfrac{1}{2}\sigma_{e0}b_1w^{-1} - \dfrac{1}{2}\sigma_{e0}b_1e^{2\iota\beta_e}w^{-3};\\ \dfrac{\psi(w)}{R} = B(w)\ln w - A\bar{b}_1 + \dfrac{\beta_1w^{-1} + \beta_2w^{-2} + \beta_3w^{-3} + \beta_5w^{-5}}{\phi'(w)},\end{cases} \tag{55}$$

with $A$ and $B(w)$ given by (42) and (47), respectively, and

$$\beta_1 = -\frac{1}{2}\sigma_{e0}\left(1 + b_1\bar{b}_1\right), \beta_2 = -A\left(1 + b_1\bar{b}_1\right), \beta_3 = -\frac{1}{2}\sigma_{e0}e^{2\iota\beta_e}\left(1 + 3b_1\bar{b}_1\right), \beta_5 = -\sigma_{e0}e^{2\iota\beta_e}b_1. \tag{56}$$

According to the equations (24) - (31) in Wang and Wang 2017, K-M potentials have

$$\begin{cases}\dfrac{\varphi(w)}{R} = A\ln w - \dfrac{1}{2}\sigma_{e0}b_1w^{-1} - \dfrac{1}{2}\sigma_{e0}b_1e^{2\iota\beta_e}w^{-3};\\ \dfrac{\psi(w)}{R} = B(w)\ln w - \dfrac{\overline{t(w)}}{t'(w)}\dfrac{\varphi'(w)}{R} - \dfrac{1}{2}\sigma_{e0}\left(w^{-1} + e^{2\iota\beta_e}w^{-3}\right).\end{cases} \tag{57}$$

It is easy to check that the equations (56) and (57) are exactly the same, but the constant part in $(57)_2$ deviates from initial conception and no clear explanation is given. In addition, based on (31) of Wang and Wang 2017, we find that the toroidal normal stress (TNS) in (62) of Wang and Wang 2017 should be corrected as $\tau_{\theta\theta} = -8\tau_{e0}\frac{2m\cos(\alpha-\theta)-(m^2+1)\cos(\alpha+\theta)}{1+m^2-2m\cos 2\theta}$ when the uniform energy flux $J_u^\infty$ is solely applied, and the maximum stress appears at $\beta = \pi/2$ instead of $\beta = 0, \pi$ (see Fig. 5 of Wang and Wang 2017). As a special case of Wang and Wang 2017, Zhang and Wang 2016 and Zhang *et al.* 2017 discuss the ellipse problem with uniaxial loading.

### 3.3 Solutions of a non-elliptical cavity and comparison

For a non-elliptical cavity, a truncated form solution is proposed by Yu *et al.* (2019)

$$\varphi(w) = RA\ln w + \sum_{k=1}^{2M} p_k w^{-k}, \psi(w) = RB(w)\ln w + \sum_{k=1}^{2M} q_k w^{-k} \tag{58}$$

where $M$ is the truncation order of the potentials, the coefficients $p_k$ and $q_k$ are determined by solving a set of linear equations obtained from the traction boundary condition (45). It is obvious that equation (58) has different structure from ours (49). We will present two non-elliptic examples for comparison.

First consider a triangular cavity characterized by $t(\eta) = R(\eta + b_2\eta^{-2}), |\eta| = 1$, the K-M potentials obtained from our method have explicit expressions

$$\begin{cases} \dfrac{\varphi(w)}{R} = A\ln w - \bar{A}b_2 w^{-1} - \dfrac{8}{3}\sigma_{e0} b_2 w^{-2} - \dfrac{2}{3}\sigma_{e0} b_2 e^{2\iota\beta_e} w^{-4}; \\ \dfrac{\psi(w)}{R} = B(w)\ln w - \bar{A}b_2\bar{b}_2 - 2\sigma_{e0}\bar{b}_2 e^{2\iota\beta_e} + \dfrac{\beta_1 w^{-1} + \beta_2 w^{-2} + \beta_3 w^{-3} + \beta_4 w^{-4} + \beta_6 w^{-6}}{\phi'(w)}. \end{cases} \tag{59}$$

with

$$\beta_1 = -\sigma_{e0}\left(\frac{1}{2} + \frac{16}{3} b_2\bar{b}_2\right), \beta_2 = -\sigma(1 + 2b_2\bar{b}_2), \beta_3 = -\bar{\sigma}b_2(1 + 2b_2\bar{b}_2) - \sigma_{e0}e^{2\iota\beta_e}\left(\frac{1}{2} + \frac{8}{3}b_2\bar{b}_2\right),$$
$$\beta_4 = -\frac{13}{3}\sigma_{e0}b_2, \beta_6 = -\frac{5}{3}\sigma_{e0}e^{2\iota\beta_e}b_2. \tag{60}$$

Form Yu *et al.* (2019), we find that when $M \geq 2$, the solutions are expressed by

$$\begin{cases} \dfrac{\varphi(w)}{R} = A\ln w - \bar{A}b_2 w^{-1} - \dfrac{8}{3}\sigma_{e0} b_2 w^{-2} - \dfrac{2}{3}\sigma_{e0} b_2 e^{2\iota\beta} w^{-4}; \\ \dfrac{\psi(w)}{R} = B(w)\ln w + \dfrac{1}{R}\sum_{k=1}^{2M} q_k w^{-k}. \end{cases} \tag{61}$$

For the first potential $(61)_1$, due to the characteristics of its finite number of terms, the complete expression of the perturbation part can be obtained when truncation order $M$ is small, while the complete expression of the second perturbation potential in $(49)_2$ cannot be exactly expressed in the truncated form $(61)_2$. We chose the maximum shear stress (MSS) $\tau_{max}$ to show the deficiency of the truncation treatment. The MSS on the boundary can be written as

$$\tau_{max} = \sqrt{\left(\frac{\sigma_{22} - \sigma_{11}}{2}\right)^2 + \tau_{12}^2}, \tag{62.1}$$

where by equation $(31)_2$, the stress components $\sigma_{11}$, $\sigma_{22}$ and $\tau_{12}$ have the form as

$$\sigma_{22} - \sigma_{11} + 2\iota\tau_{12} = 2\sigma_{e0}\frac{(e^{-\iota\beta_e} - e^{\iota\beta_e}\eta^{-2})}{\phi'(\eta)}\left[e^{\iota\beta_e}\left(\frac{1}{2}\eta^{-2} + \bar{b}_1\ln\eta + \sum_{k=2}^{N}\frac{k}{k-1}\bar{b}_k\eta^{k-1}\right)\right.$$
$$\left. + e^{-\iota\beta_e}\left(-\ln\eta + \sum_{k=1}^{N}\frac{k}{k+1}\bar{b}_k\eta^{k+1}\right)\right] + 2[\bar{t}\varphi''(t) + \psi'(t)]. \tag{62.2}$$

The MMSs for the case with $b_2 = 1/4$ (Fig. 2a) and $\beta_e = \beta_u = \pi/2$ are shown in Fig. 3, it can be seen that the higher the truncation order $M$, the closer Yu's result to our analytical solution. When $M = 20$, the maximum error between Yu's result and ours is $0.02780\sigma_{e0}$ for the case of $J_e^\infty$ applied alone, and the maximum error between Yu's result and ours is $0.01387\sigma_{u0}$ for the case of $J_u^\infty$ applied alone.

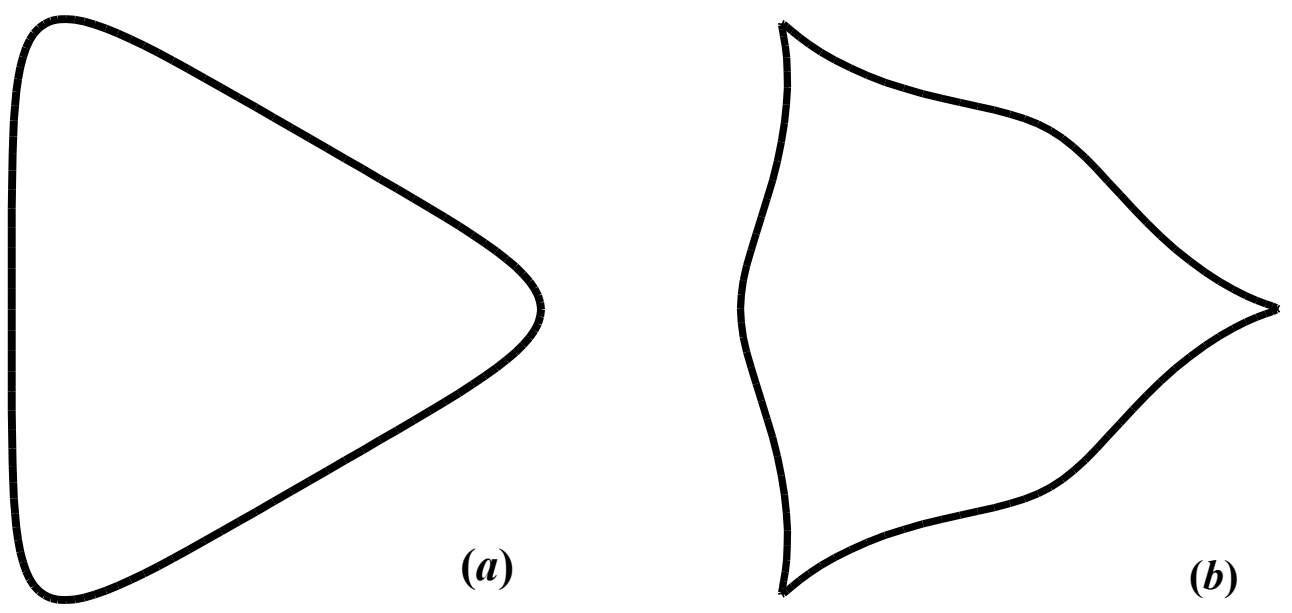

**Figure 2.** Cavities described by Laurent polynomials: (*a*). $t(\eta) = R(\eta + b_2\eta^{-2}), |\eta| = 1$; (*b*). $t(\eta) = R(\eta + 1/4\eta^{-2} + 1/12\eta^{-5}), |\eta| = 1$.

Furthermore, if one more coefficient is involved so that $t(\eta) = R(\eta + 1/4\eta^{-2} + 1/12\eta^{-5}), |\eta| = 1$, as shown in Fig. 2b, the comparison of MMSs is presented in Fig. 4. It is obvious that even a much higher truncation order *M* cannot meet the accuracy requirement when one more term is added in the Laurent polynomial, especially around the tip.

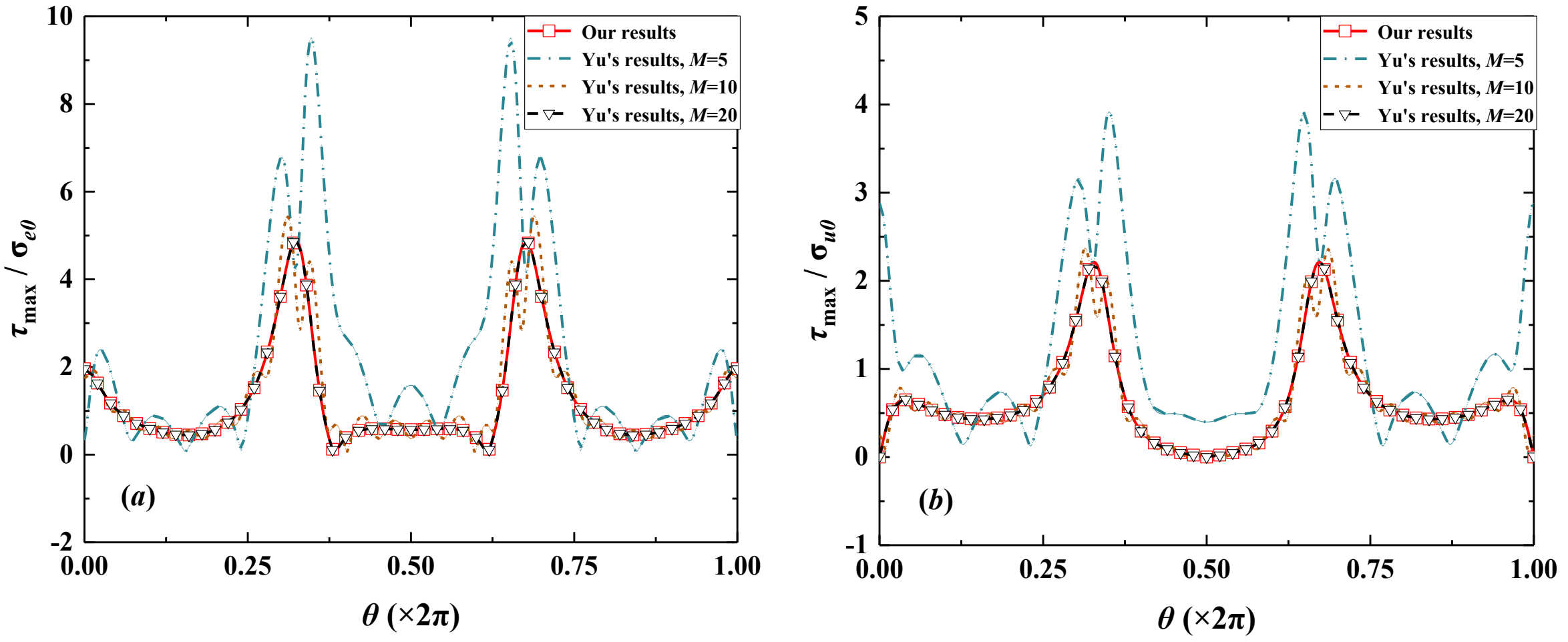

**Figure 3.** MSS scaled by $\sigma_{e0}$ or $\sigma_{u0}$ along the boundary of the cavity $t(\eta) = R(\eta + 1/4\eta^{-2}), |\eta| = 1$ with $\beta_e = \beta_u = \pi/2$: (*a*). $J_e^\infty$ solely applied; (*b*). $J_u^\infty$ solely applied.

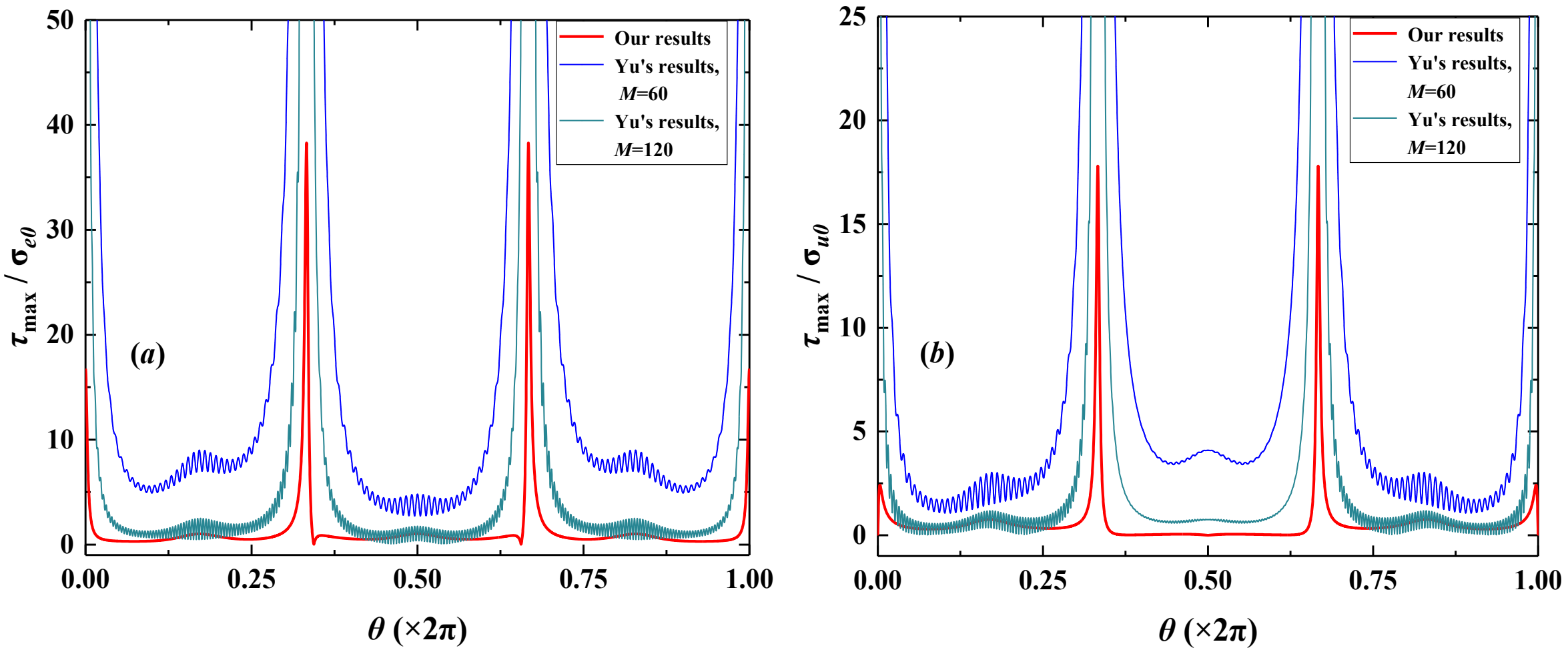

**Figure 4.** MSSs scaled by $\sigma_{e0}$ or $\sigma_{u0}$ along the boundary of the cavity $t(\eta) = R(\eta + 1/4\eta^{-2} + 1/12\eta^{-5}), |\eta| = 1$ with $\beta_e = \beta_u = \pi/2$: (*a*). $J_e^\infty$ solely applied; (*b*). $J_u^\infty$ solely applied.

## 4 Illustrations and analyses of thermoelectric and stress concentration

In comparison with elliptic cavity, few studies are related to polygonal cavities, let alone the tip patterns. In this section, we are concerned with the thermoelectric and stress concentration around the tip of the cavity. In the following, the energy flux and electric current density are scaled by $J_e^\infty$ and $J_u^\infty$ respectively, the toroidal normal stress (TNS) scaled by $\sigma_{e0}$ or $\sigma_{u0}$. We use triangle, square and pentagram as typical shapes, whose mapping functions are expressed by (Savin, 1961; Zou and He 2018):

$$t(\eta) = R\left(\eta + \frac{1}{3}\eta^{-2} + \frac{1}{45}\eta^{-5} + \frac{1}{162}\eta^{-8}\right), R = 0.64087, \tag{63.1}$$

$$t(\eta) = R\left(\eta + \frac{1}{6}\eta^{-3} + \frac{1}{56}\eta^{-7} + \frac{1}{176}\eta^{-11}\right), R = 0.59011, \tag{63.2}$$

$$\begin{aligned} t(\eta) = R\Big(&\eta + \frac{3}{10}\eta^{-4} - \frac{13}{225}\eta^{-9} + \frac{4}{125}\eta^{-14} - \frac{214}{11875}\eta^{-19} + \frac{1231}{93750}\eta^{-24} \\ &- \frac{20974}{2265625}\eta^{-29} + \frac{2908}{390625}\eta^{-34} - \frac{441199}{76171875}\eta^{-39} + \frac{95763}{19531250}\eta^{-44} \\ &- \frac{6890609}{1708984375}\eta^{-49}\Big), R = 0.74224, \end{aligned} \tag{63.3}$$

respectively, where $|\eta| = 1$, the size parameters $R$ are chosen to ensure that the different shapes of cavities have the same area in order to carry out the comparison. The shapes and tip indices of cavities expressed by (63) are drawn in Fig. 5*a*-*c*, respectively.

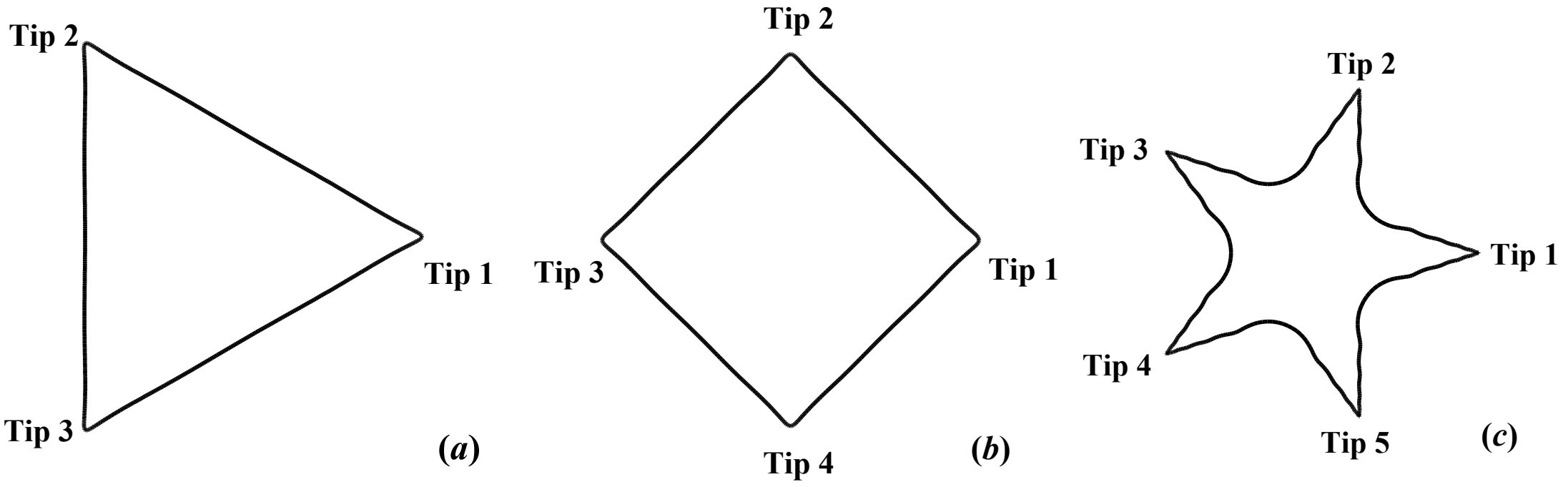

**Figure 5.** Three cavities described by Laurent polynomials: (*a*). triangle, (*b*). square, (*c*). pentagram.

### 4.1 Current density, energy flux, and TNS distribution along cavity contour

Curvature is a key parameter to describe the characteristics of the shape, and its formula is obtained in Zou and He (2018):

$$K_0 = \frac{1}{R|\phi'(\eta)|}\mathrm{Re}\left[1 + \frac{\eta\phi''(\eta)}{\phi'(\eta)}\right]. \tag{64}$$

The curvatures of the three shapes are shown in Fig. 6. We can find that the maximum curvature are pentagram $>$ triangle $>$ square. It is worth noting that each tip of the pentagram shows bimodal characteristics, which means that the concentration may occur in both peaks.

According to formulas (13) and (14), and combined with equations (21) and (26), the components of electric current density and energy flux have relations (Zhang and Wang 2016)

$$J_n^e - iJ_\tau^e = \frac{\eta\phi'(\eta)}{|\phi'(\eta)|}[J_1^e - iJ_2^e], \qquad J_n^u - iJ_\tau^u = \frac{\eta\phi'(\eta)}{|\phi'(\eta)|}[J_1^u - iJ_2^u], \tag{65}$$

where $J_\tau^*$ and $J_n^*$ are the toroidal direction and radial direction of the boundary $\Gamma$. Due to the insulation of electrical and energy flux at the boundary of cavity, namely $J_n^e = J_n^u = 0$, the toroidal electric current density $J_\tau^e$ and toroidal energy flux $J_\tau^u$ have

$$J_\tau^e = J_e^\infty \mathrm{Im}\left[\frac{e^{\iota\beta_e}\eta^{-1} - e^{-\iota\beta_e}\eta}{\left|\overline{\phi'(\eta)}\right|}\right], \qquad J_\tau^u = J_u^\infty \mathrm{Im}\left[\frac{e^{\iota\beta_e}\eta^{-1} - e^{-\iota\beta_e}\eta}{\left|\overline{\phi'(\eta)}\right|}\right]. \tag{66}$$

From (66), we can find that $J_\tau^e$ and $J_\tau^u$ on the boundary of cavity have the same form. From Fig. 6, the distribution of $J_\tau^e$ $(J_\tau^u)$ under the loading direction $\beta = 0$ is shown, it is obvious that $J_\tau^e$ $(J_\tau^u)$ have the same symmetry on the boundary as the symmetry of the shape and the loading.

Since the boundary of the cavity is traction-free, implying radial normal stress $\sigma_n = 0$, and based on the relation $\sigma_\tau + \sigma_n = \sigma_{11} + \sigma_{22}$ (Savin, 1961), the TNS $\sigma_\tau$ on the boundary is written as

$$\sigma_\tau = 4\mathrm{Re}[\varphi'(t)] + \frac{\lambda\sigma}{4k} f(t)\overline{f(t)} = 4\mathrm{Re}\left[\frac{\varphi'(\eta)}{t'(\eta)}\right] + 2\sigma_{e0}\left(2 + e^{-2\iota\beta_e}\eta^2 + e^{2\iota\beta_e}\eta^{-2}\right), \eta = e^{\iota\theta}. \tag{67}$$

Similarly, stress is symmetrically distributed on three symmetrical polygons, as shown in Fig. 6. Because all tips with different orientations have the same features for the cavities expressed by (63), we will focus on tip 1 of cavity in the following analyses.

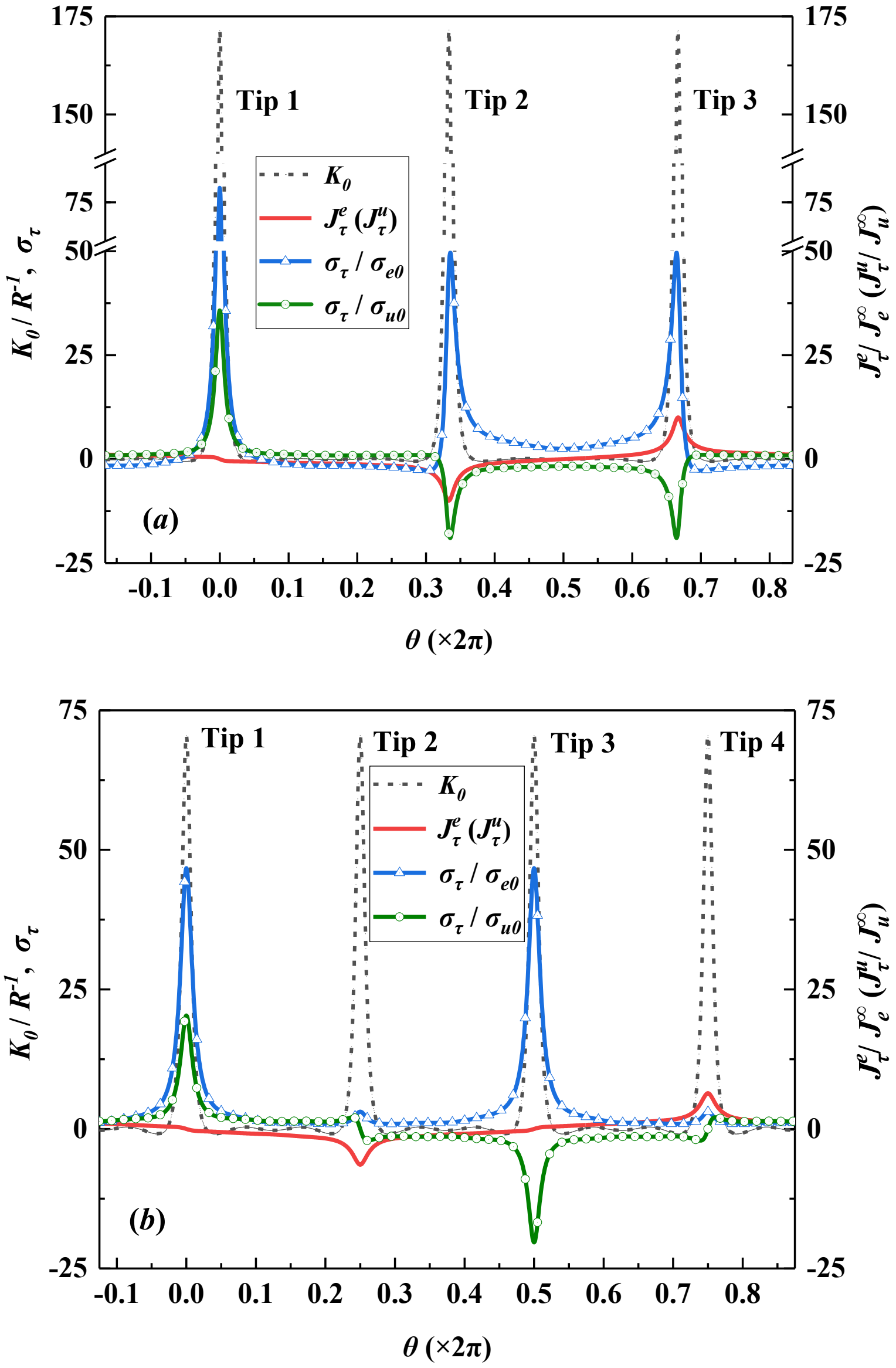

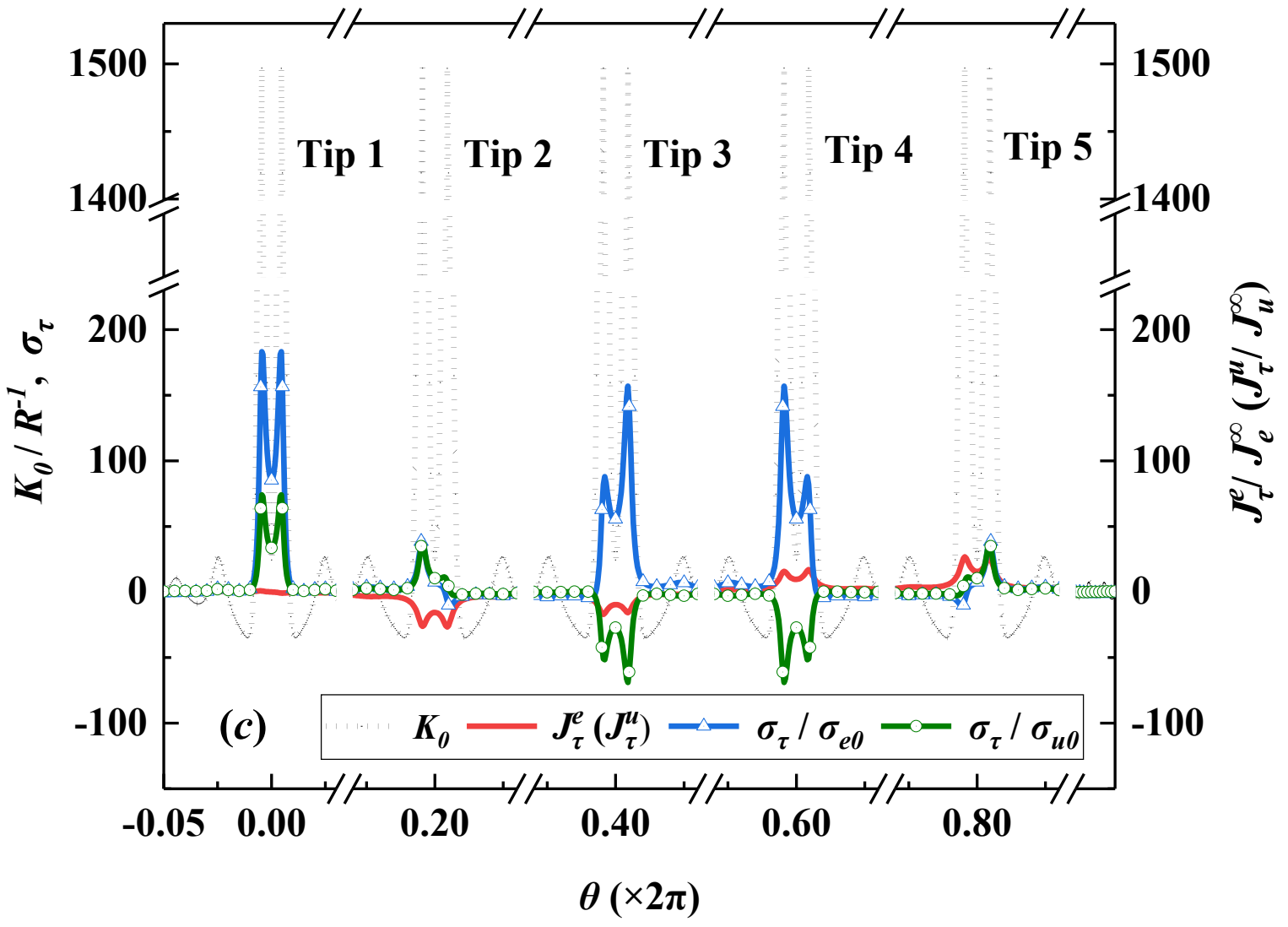

**Fig. 6.** Curvature scaled by $R^{-1}$, toroidal normal stress scaled by $\sigma_{e0}$ or $\sigma_{u0}$, toroidal electric current density (toroidal energy flux) scaled by $J_e^\infty$ ($J_u^\infty$) along the contour of the cavity with $\beta = 0$: (*a*) triangle, (*b*) square, (*c*) pentagram.

## 4.2 Change of toroidal electric current density (toroidal energy flux) around the tip

The toroidal electric current densities (toroidal energy fluxes) at tips of the cavities are presented in Fig. 7. For the cases of $\beta = 0$ and $\pi$, there are smaller thermoelectric concentrations for all three shapes, and $J_\tau^e = J_\tau^u = 0$ at $\theta = 0$. For other cases of $\beta \neq 0, \pi$, the $J_\tau^e$ ($J_\tau^u$) around the tip of triangle and square have the same distribution trends that are different from that of pentagram, and due to the bimodal characteristics shown in Fig. 6c, the $J_\tau^e$ ($J_\tau^u$) obvious changes at the two peaks of pentagram tip. Under the same direction of $J_e^\infty$ ($J_u^\infty$), the maximums $J_\tau^e$ ($J_\tau^u$) obtained around the maximum curvature point are positively correlated with the maximum curvature for three shapes, namely pentagram > triangle > square. Curvature can be used as a reference for degree of thermoelectric concentration.

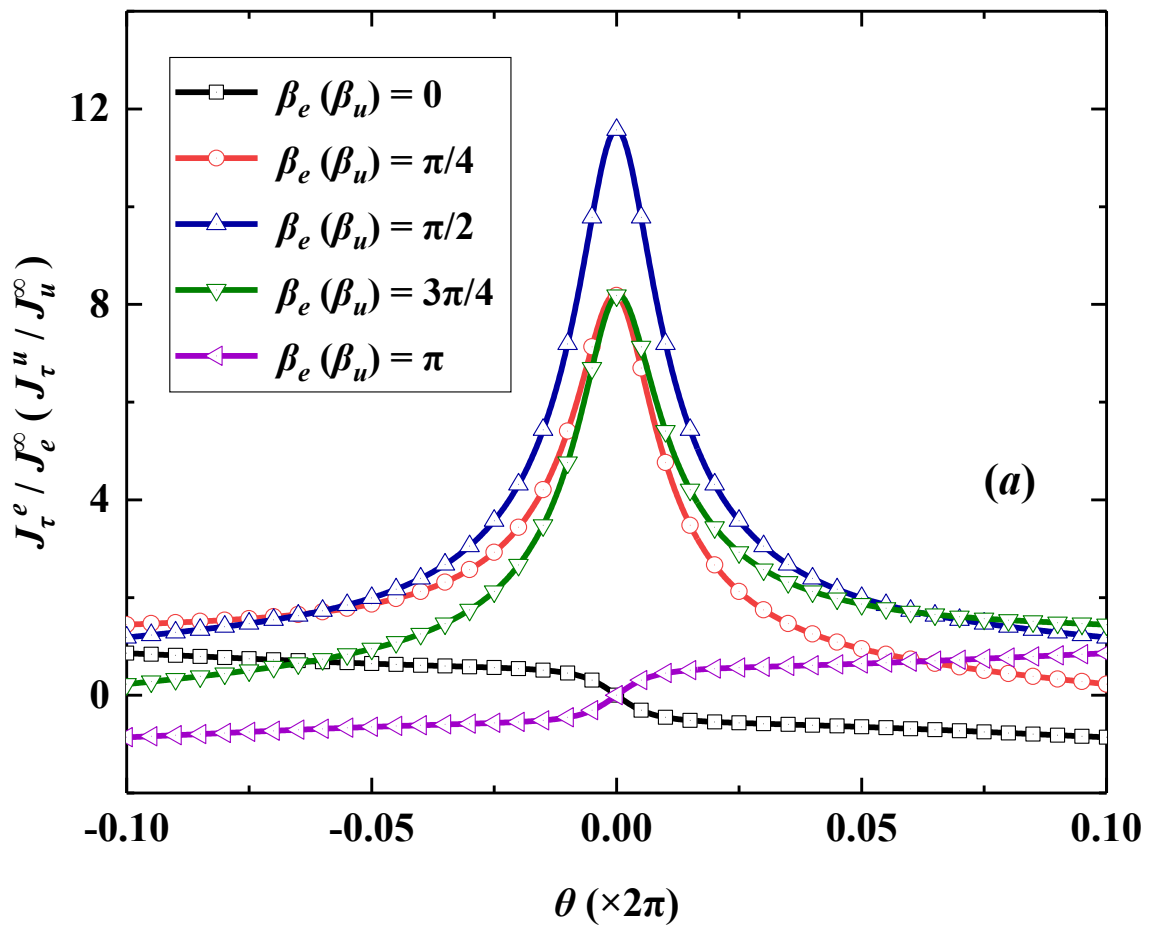

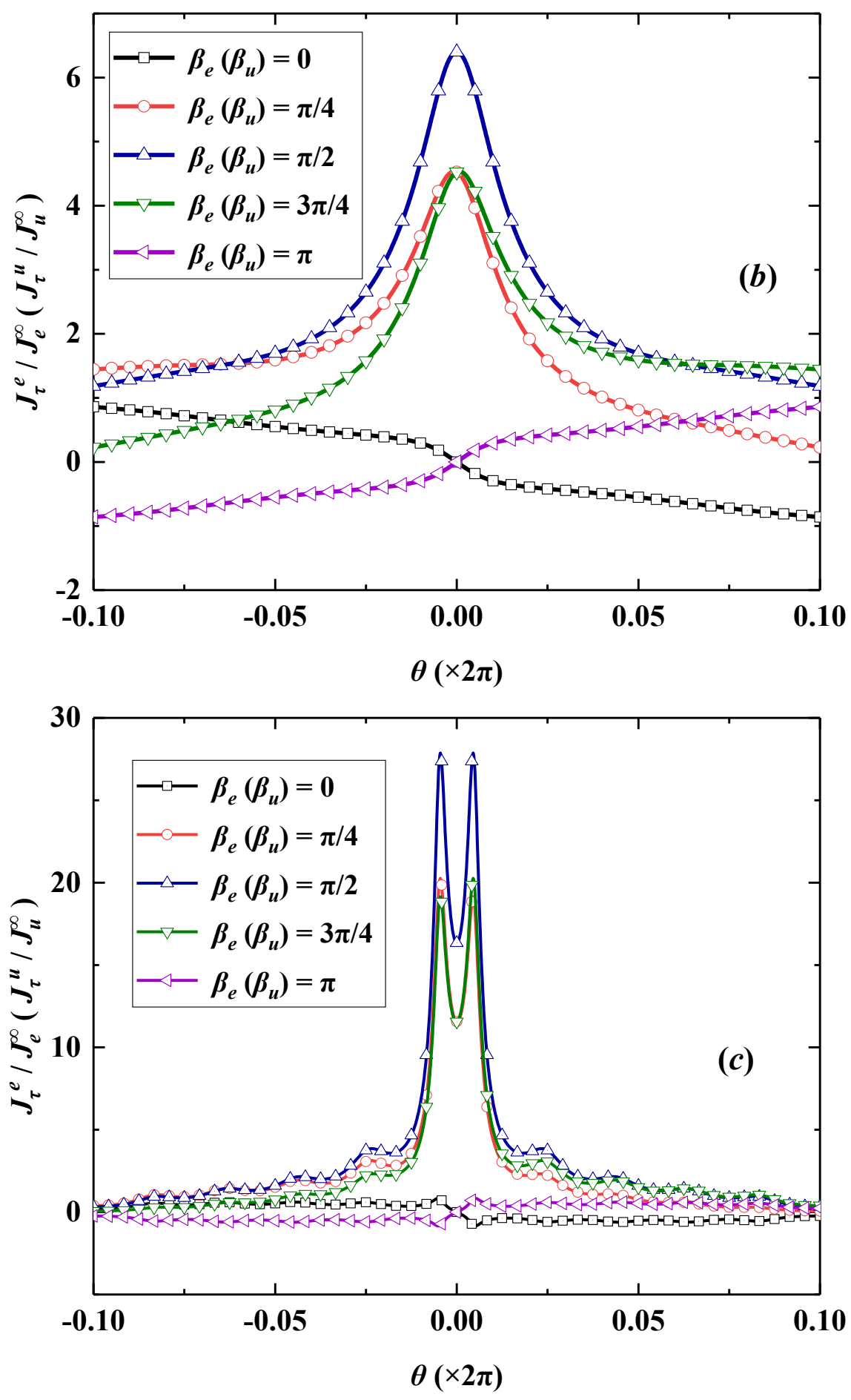

**Figure 7.** Toroidal electric current densities (energy fluxes) are scaled by $J_e^\infty$ or $J_u^\infty$ along the tip of the cavities: (*a*) triangle, (*b*) square, (*c*) pentagram.

### 4.3 Change of TNS around the tip

The TNSs at the tips are shown in Fig. 8. It is clear that the distribution of stress is also closely related to the curvature of the tip for three different shapes. In the cases where $J_u^\infty$ solely applied, the stress is null at $\theta = 0$ when $\beta = 0$. In the other cases, the TNSs increase to varying degrees arise from the surge of curvature around the tips, and the maximum stress is also achieved around the maximum curvature point. Due to the curvature at the tip of pentagram exhibits bimodal characteristics, the maximum stress is obtained simultaneously at the two peaks when the direction of loading is $\beta = \pi/2$, which means that when the material is damaged, the crack may expand in both directions. Similarly, the maximum stress depends on the maximum curvature under the same load direction, that is, pentagram $>$ triangle $>$ square. Based on the above, large curvature is often accompanied by strong stress for polygonal cavities, curvature is also expected to be used as a tool to judge stress concentration in engineering applications.

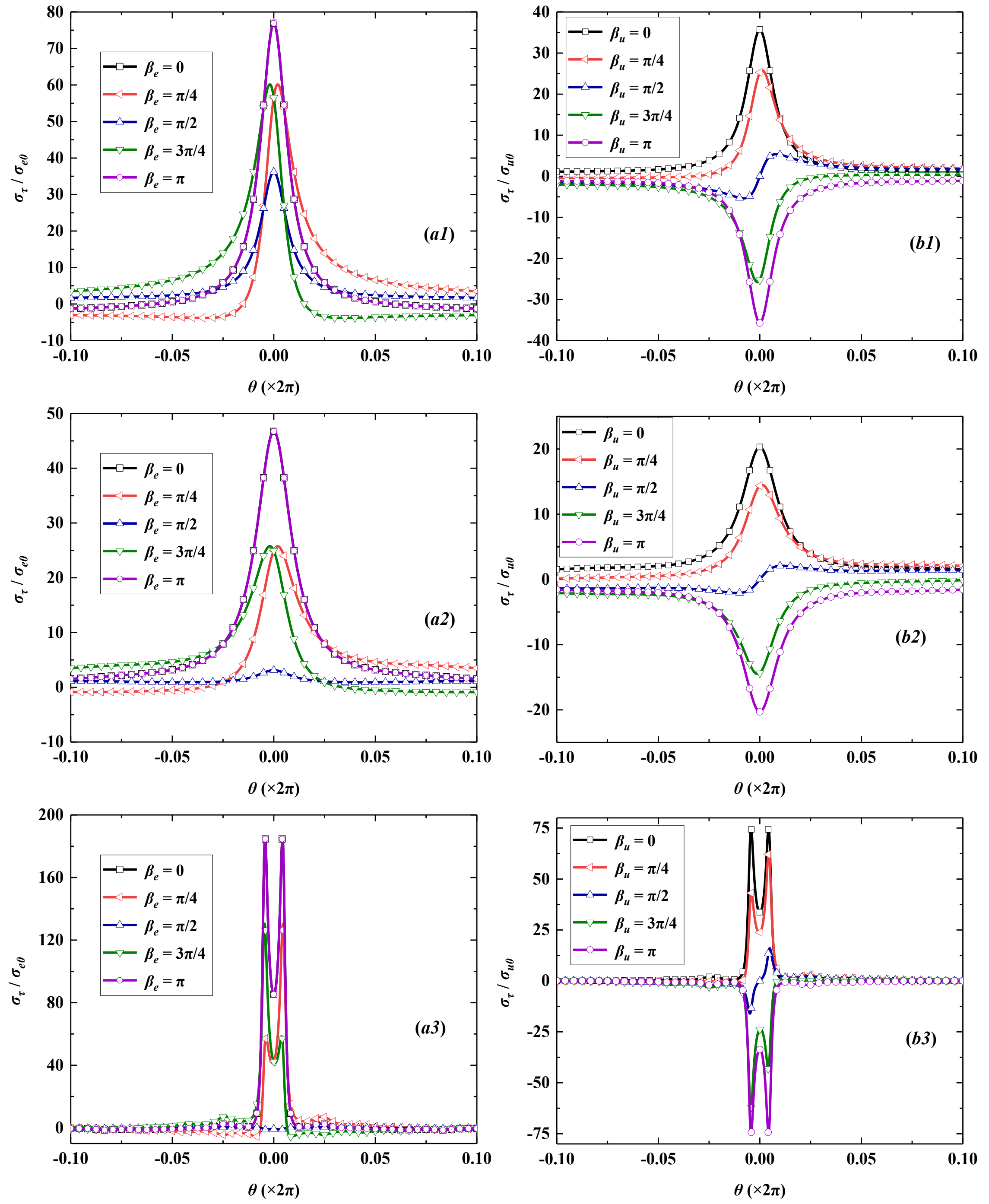

**Figure 8.** Toroidal normal stress (TNS) scaled by $\sigma_{e0}$ or $\sigma_{u0}$ along the tips of different cavities: (*a1*), (*b1*) triangle; (*a2*), (*b2*) square; (*a3*), (*b3*) pentagram.

### 4.4 Thermoelectric and stress concentration at the tip

We consider the variation of maximum thermoelectric concentration $|J_\tau^e|_{max}$ ($|J_\tau^u|_{max}$), and maximum stresses concentration $|\sigma_\tau|_{max}$ under different loading directions for the three shapes described by equation (63). It should be noted that the absolute value is used to filter the maximum value and the contour of $\theta \in (-\pi/5, \pi/5)$ around the tip is selected as the calculation interval.

The distribution of $|J_\tau^e|_{max}$ ($|J_\tau^u|_{max}$) is shown in Fig. 9. We can found that the maximum value appears at $\beta = \pi/2$ and the minimum value appears at $\beta = 0, \pi$ for three shapes, that is, when the loading direction is perpendicular to the symmetry axis of the tip, the strongest thermoelectric concentration can be utilized, while when the loading direction is parallel to the symmetry axis of the tip, there is the minimum thermoelectric concentration. This can make better use of cavities to improve thermoelectric performance in engineering.

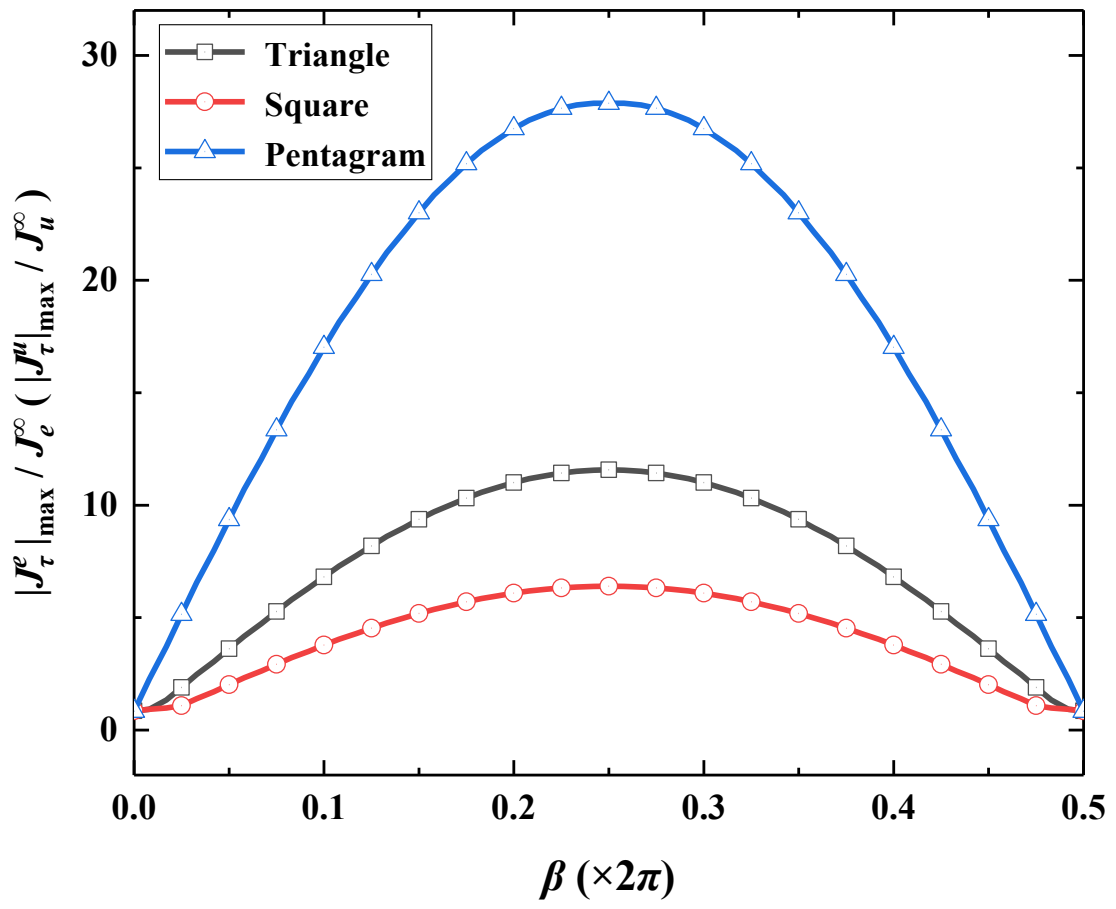

**Figure 9.** Concentration of current density (energy flux) under different loading directions.

The stress concentration $|\sigma_\tau|_{max}$ at the tips of the three shapes is given in Fig. 10. For triangle and square, the maximum stress concentration occurs when the loading is parallel to the symmetry axis of the tip ($\beta = 0, \pi$), while the minimum stress concentration occurs when the load is perpendicular to the symmetry axis of the tip ($\beta = \pi/2$). For the case of pentagram, the minimum stress concentration occurs when $\beta = \pi/2$, the maximum stress concentration occurs when the loading direction about $\beta = 1/30\ (12°)$ away from the symmetry axis of the tip caused by the platform between the two peaks (the minimum stress can be occur at the platform between the two peaks, while the maximum concentration cannot). Based on the above discussion, the damage caused by maximum stress can be reduced to improve the service life of thermoelectric materials in engineering.

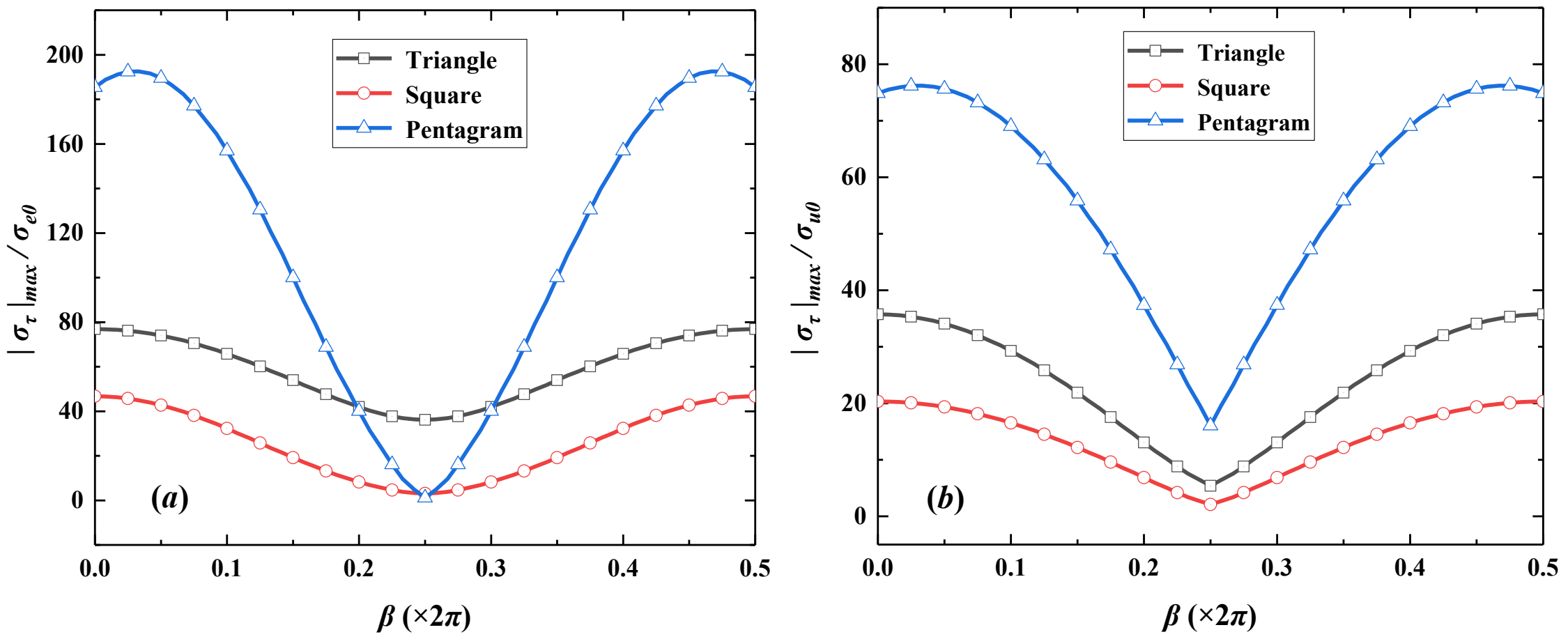

**Figure 10.** Concentration of TNSs scaled by $\sigma_{e0}$ or $\sigma_{u0}$ along the tip under different load directions: ($a$). $J_e^\infty$ solely applied; ($b$). $J_u^\infty$ solely applied.

According to the discussion in Section 5.1 and Section 5.2, the maximum concentration occurs around the maximum curvature point. Thermoelectric $|J_\tau^e|$ ($|J_\tau^u|$) and stress $|\sigma_\tau|$ changes at the maximum curvature point are obtained in Fig. 11-12, it should be pointed out that we chose one of the peaks at tip 1 of the pentagram cavity to calculate. For pentagram with bimodal characteristics, comparing Fig. 9 and Fig. 10, it is obvious that the value of the maximum curvature point is not the maximum value at the tip, which is caused by the alternating appearance of the maximum value point at the two peaks of the tip and the maximum value point moves around the maximum curvature point, we can see that the minimum $|\sigma_\tau|$ occurs when the load direction is $\beta = 2/\pi$, but the maximum $|\sigma_\tau|$ occurs when the load direction is $\beta = 0, \pi$ for the maximum curvature point. For triangle and square with symmetrical tips, they are very similar to Fig. 9 and Fig. 10 both in terms of values and distribution trends.

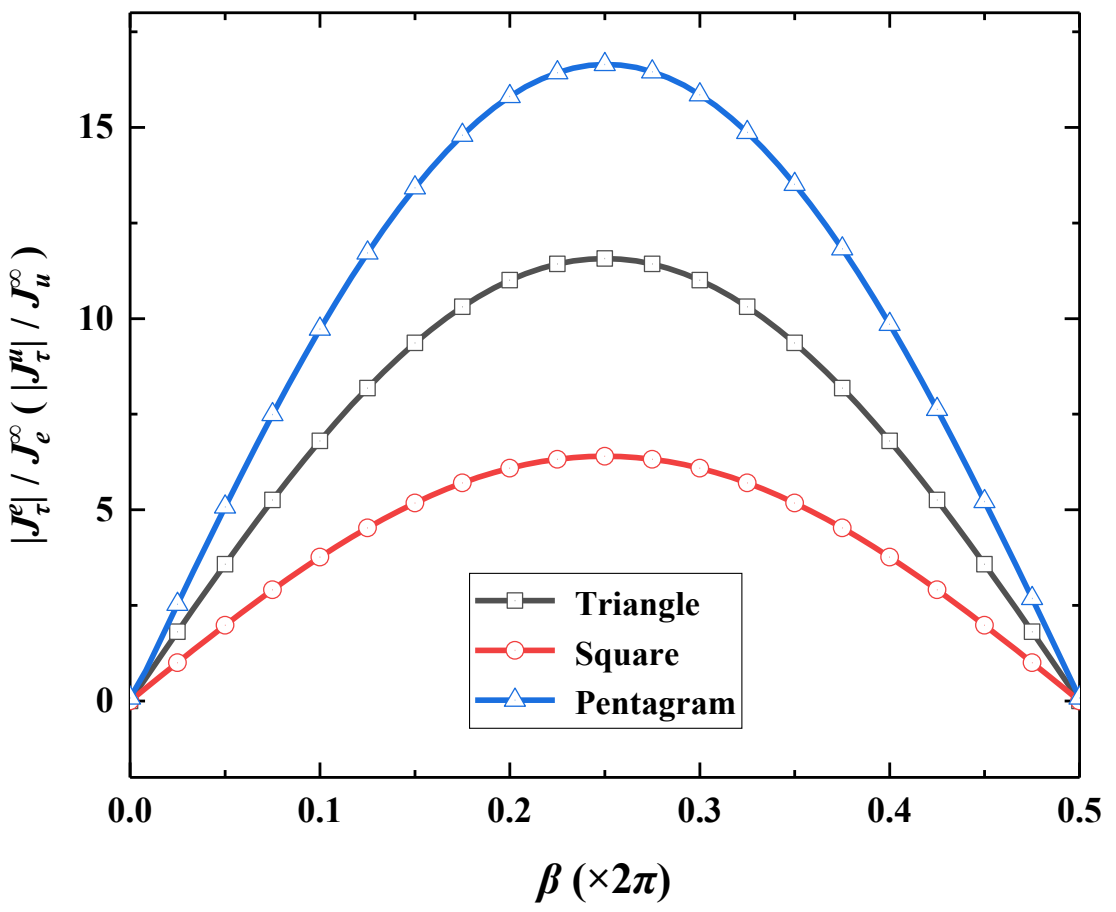

**Figure 11.** Ccurrent density (energy flux) at the maximum curvature point under different loading directions.

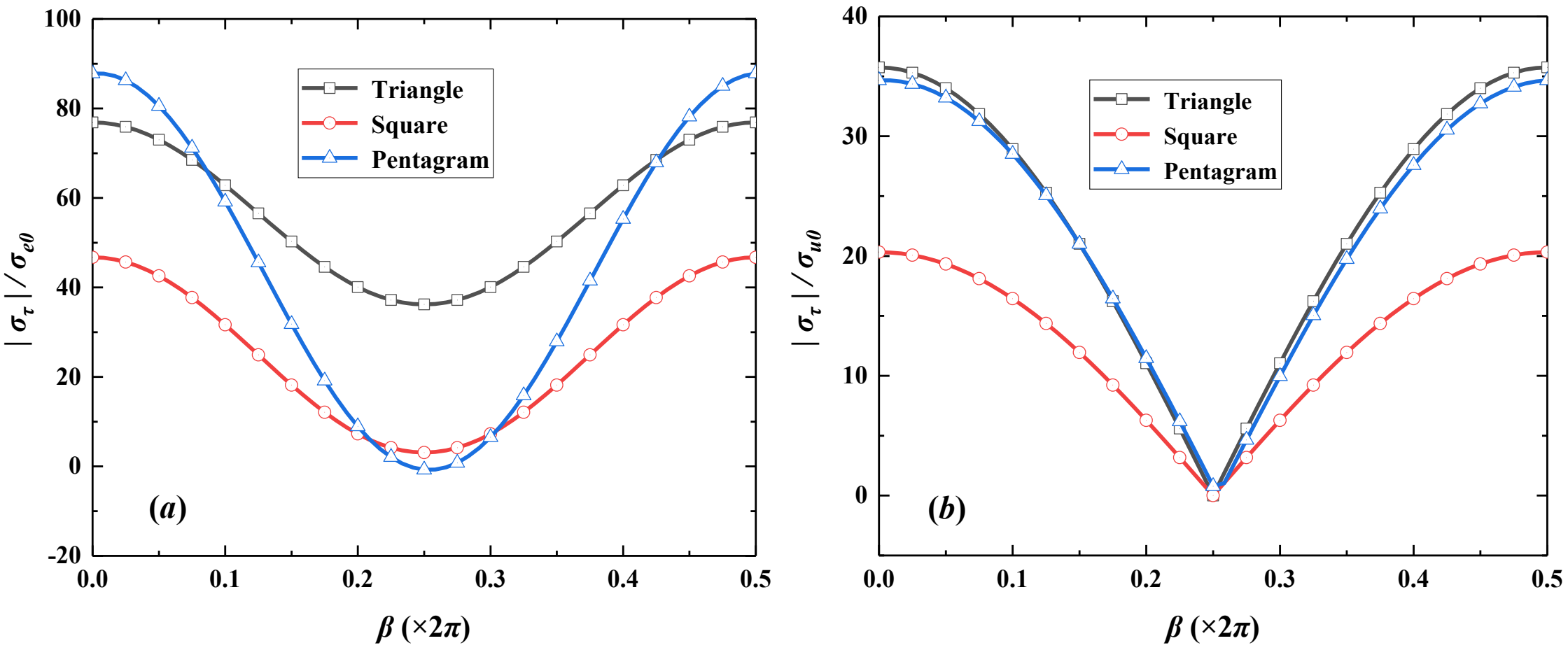

**Figure 12.** TNS scaled by $\sigma_{e0}$ or $\sigma_{u0}$ at the maximum curvature point under different loading directions: (*a*). $J_e^\infty$ solely applied; (*b*). $J_u^\infty$ solely applied.

Similar to the pentagram shown above, to further discuss the difference between the value of the maximum curvature point and the maximum value on the tip of triangle and square, the thermoelectric difference $|\Delta_\tau^e|$ $(|\Delta_\tau^e|)$ and stress difference $|\Delta_\tau|$ between the two values are shown in Fig. 13. It can be seen that $|\Delta_\tau^e|(|\Delta_\tau^e|) = 0$ when $\beta = \pi/2$ from Fig. 13 (*a*), it means that when the load direction is perpendicular to the tip axis of symmetry, the maximum thermoelectric concentration occurs at the maximum curvature point. From Fig. 13 (*b*), the maximum stress concentration occurs at the maximum curvature point at $\beta = 0, \pi/2, \pi$ for the case of $J_e^\infty$ is solely applied and at $\beta = 0, \pi$ for the case of $J_u^\infty$ is solely applied.

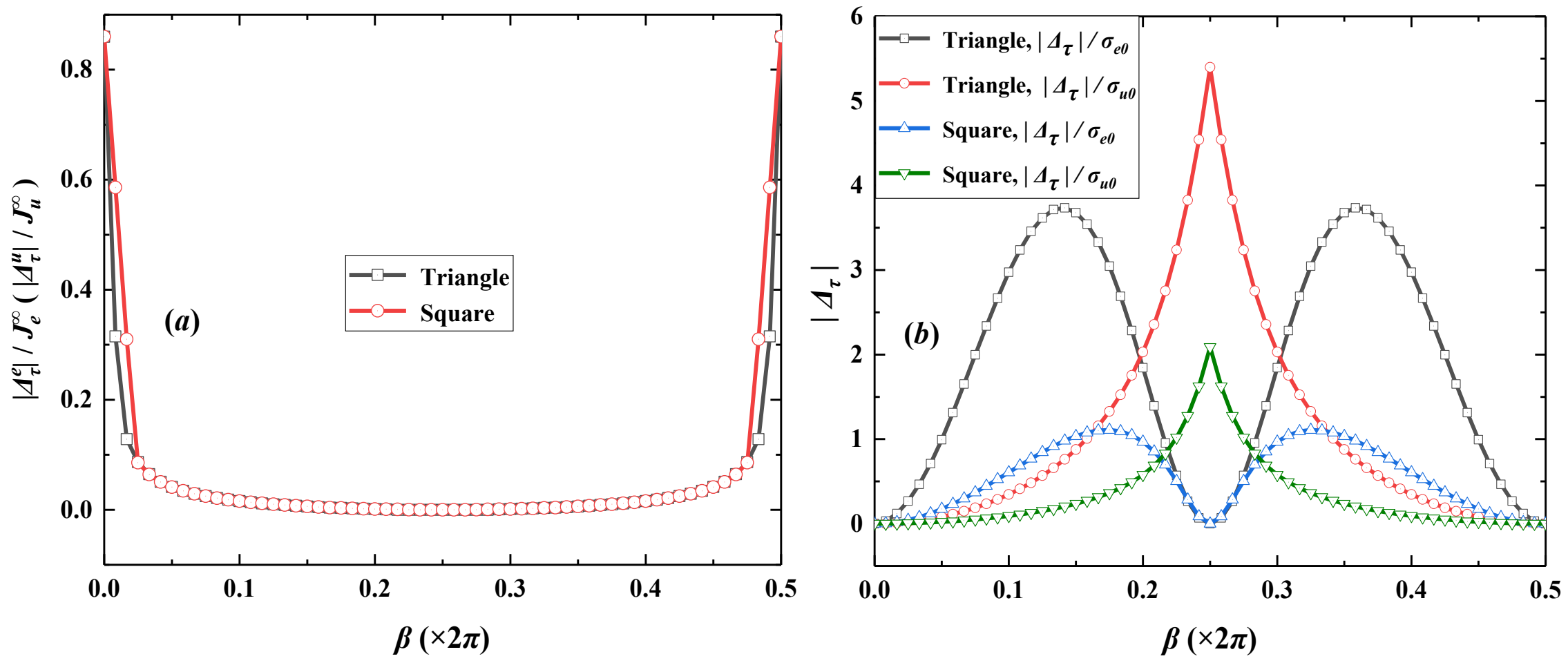

**Figure 13.** Value difference between the maximum curvature point and the maximum value point at the tip under different loading directions: (*a*). current density (energy flux); (*b*). TNS scaled by $\sigma_{e0}$ or $\sigma_{u0}$.

## 5 Conclusion

In this paper, a hot problem, the two-dimensional isotropic thermoelectric material containing a smooth cavity with an electric insulated and adiabatic boundary, and subjected to uniform electric current density or uniform energy flux at infinity, is studied by the K-M potential theory. The explicit analytic solutions together with the rigid-body translation are first derived when the cavity is characterized by the Laurent polynomial.

Triangle, square and pentagram are considered as representative shapes of the cavity to study the concentration under the remote loadings along different directions. We find that the electric current density (energy flux) and stress enhanced with the surge of curvature around the tip of the cavity, and it is positively correlated with the maximal curvature for three polygons with the same area, that is, pentagram > triangle > square. The maximum thermoelectric and stress concentration reaches the maximum or minimum when the loading direction is parallel to or perpendicular to the symmetry axis of the tip, and this is not the same as the pentagram cavity with bimodal characteristics. The maximum thermoelectric and stress concentration occur the maximum curvature point only under special loading direction, and in most cases near the maximum curvature point.

## Appendix A. Detailed derivations of explicit analytical solutions

Following the analysis of Zou and He (2018), the boundary equation (48) with respect to the perturbation potentials $\varphi_p(\eta)$ and $\psi_p(\eta)$ is reasonably expanded as a power series in terms of $\eta$, and according to the series form (37), the perturbation potentials and rigid-body translation can be obtained by analyzing non-positive power series. Thus, an equivalent relation "$\sim$" can be introduced to connect the expressions having the same non-positive power.

First of all, for the purposes of further work, the function of non-positive powers is defined as

$$G(\eta^{-1}) \equiv \sum_{k=0}^{N} g_k \eta^{-k} \sim \frac{\phi(\eta)}{\overline{\phi'(\eta)}} = \left(1 - \sum_{l=1}^{N} l\bar{b}_l \eta^{l+1}\right)^{-1} \sum_{k=1}^{N} b_k \eta^{-k}, \tag{A.1}$$

we can give the equivalent relation

$$\sum_{k=0}^{N} g_k \eta^{-k} \sim \sum_{k=0}^{N-2} \left(\sum_{l=1}^{N-k-1} l g_{k+l+1} \bar{b}_l\right) \eta^{-k} + \sum_{k=1}^{N} b_k \eta^{-k}, \tag{A.2}$$

then $\{g_k, k = 1, \cdots, N\}$ are determined by $b_k$ as (50) besides

$$g_0 = \sum_{l=1}^{N-1} l g_{l+1} \bar{b}_l. \tag{A.3}$$

According to (A.2), we have

$$\frac{t(\eta)}{\overline{t'(\eta)}} \overline{\varphi_p'(\eta)} \sim -R \sum_{n=2}^{N} g_k \eta^{-k} \sum_{l=1}^{\infty} l\bar{\alpha}_l \eta^{l+1} \sim -R \sum_{n=0}^{N-2} \eta^{-n} \sum_{l=1}^{N-n-1} l\bar{\alpha}_l g_{n+l+1}. \tag{A.4}$$

and the non-positive part of the right part of (48) is written as

$$\sum_{k=0}^{N+2} c_k \eta^{-k} = -2\mu \frac{U_0}{R} - \bar{A} \sum_{k=1}^{N} g_k \eta^{-k+1} - \sigma_{e0} \left(\frac{1}{2} b_1 \eta^{-1} + \sum_{k=2}^{N} \frac{k e^{-2\iota\beta_e}}{k-1} b_k \eta^{-k+2} + \sum_{k=2}^{N} \frac{2k^2}{k^2-1} b_k \eta^{-k} + \sum_{k=1}^{N} \frac{k e^{2\iota\beta_e}}{k+1} b_k \eta^{-k-2}\right), \tag{A.5}$$

namely

$$c_0 = -\bar{A} g_1 - \frac{2\mu U_0}{R} - 2 b_2 \sigma_{e0} e^{-2\iota\beta_e}, \tag{A.6}$$

besides (51).

Based on the above analysis, the boundary equation (48) can be balanced when the potential $\varphi_p(\eta)$ has maximal negative power $N+2$ and

$$\sum_{n=1}^{N+2} \alpha_n \eta^{-n} - \sum_{n=0}^{N-2} \eta^{-n} \sum_{l=1}^{N-n-1} l\bar{\alpha}_l g_{n+l+1} = \sum_{n=0}^{N+2} c_n \eta^{-n}. \tag{A.7}$$

It is easy to list the linear equations of the undetermined coefficients $\{\alpha_k, k = 1, \cdots, N-1\}$ as (52). An iterative method to solve (52) was proposed in Zou and He (2018), that is, starting with

$$\alpha_k^{(0)} = c_k, k = 1, 2, \ldots, N+2, \tag{A.8}$$

and doing

$$\alpha_k^{(n+1)} = c_k + \sum_{l=1}^{N-k-1} l g_{k+l+1} \bar{\alpha}_l^{(n)}, \qquad n = 0, 1, 2, \ldots \tag{A.9}$$

to achieve the convergence of the results. Then, from the balance of constant terms in (A.7), substitution of $c_0$ in (A.6) yields the formula of the rigid-body translation $U_0$ as (53).

For the potential $\psi_p(\eta)$, we can conjugate the boundary equation (48) and multiply it by $\phi'(\eta)$ to get

$$\phi'(\eta)\overline{\varphi_p(\eta)} + \overline{\phi(\eta)}\varphi_p'(\eta) + \phi'(\eta)\psi_p(\eta) = -2\mu\bar{U}_0\phi'(\eta) - RA\overline{\phi(\eta)}\eta^{-1} - R\sigma_{e0}\left(\frac{1}{2}e^{2\iota\beta_e}\eta^{-3} + \frac{1}{2}\eta^{-1} + \frac{1}{2}\bar{b}_1\eta + \sum_{k=2}^{N} \frac{k e^{2\iota\beta_e}}{k-1}\bar{b}_k\eta^{k-2} + \sum_{k=2}^{N} \frac{2k^2}{k^2-1}\bar{b}_k\eta^{k} + \sum_{k=1}^{N} \frac{k e^{-2\iota\beta_e}}{k+1}\bar{b}_k\eta^{k+2}\right)\phi'(\eta). \tag{A.10}$$

By substituting the equivalence relations

$$\phi'(\eta)\overline{\varphi_p(\eta)} \sim -R\sum_{k=1}^{N+1} k b_k \bar{\alpha}_{k+1} - R\sum_{l=1}^{N+2}\left(\sum_{k=l}^{N+2} k b_k \bar{\alpha}_{k-l+1}\right)\eta^{-l},$$

$$\overline{\phi(\eta)}\varphi_p'(\eta) \sim \eta^{-1}\varphi_p'(\eta) - R\sum_{k=1}^{N-1} k\alpha_k \bar{b}_{k+1} - R\sum_{l=1}^{N+2}\left(\sum_{k=l}^{N+2} k\alpha_k \bar{b}_{k-l+1}\right)\eta^{-l},$$

$$\phi'(\eta)\sum_{k=2}^{N} \frac{2k^2}{k^2-1}\bar{b}_k\eta^k \sim -\sum_{k=1}^{N-1} \frac{2(k+1)^2}{(k+1)^2-1} k b_k \bar{b}_{k+1} - \sum_{k=1}^{N-1}\left(\sum_{p=k+1}^{N} \frac{2(p-k+1)^2}{(p-k+1)^2-1} p b_p \bar{b}_{p-k+1}\right)\eta^{-k},$$

$$\phi'(\eta)\sum_{k=1}^{N} \frac{k}{k+1}\bar{b}_k\eta^{k+2} \sim -\sum_{k=1}^{N} \frac{k-1}{k} k b_k \bar{b}_{k-1} - \sum_{k=1}^{N-2}\left(\sum_{p=k+2}^{N} \frac{p-k-1}{p-k} p b_p \bar{b}_{p-k-1}\right)\eta^{-k},$$

$$\phi'(\eta)\sum_{k=3}^{N} \frac{k}{k-1}\bar{b}_k\eta^{k-2} \sim -\sum_{k=1}^{N-3} \frac{k+3}{k+2} k b_k \bar{b}_{k+3} - \sum_{k=1}^{N}\left(\sum_{p=1}^{N+k-3} \frac{p-k+3}{p-k+2} p b_p \bar{b}_{p-k+3}\right)\eta^{-k},$$

into (A.10), we can judge that the maximal negative power of $\phi'(\eta)\psi_p(\eta)$ must be $N+4$, and sorting out the terms of negative power have

$$\phi'(\eta)\frac{\psi_p}{R}=\sum_{k=1}^{N+2}\lambda_k\eta^{-k}-\eta^{-1}\varphi_p'(\eta)-A\eta^{-2}+2\mu\frac{\bar{U}_0}{R}[1-\phi'(\eta)]-\sigma_{e0}\left[\frac{1}{2}e^{2\iota\beta_e}\eta^{-3}+\frac{1}{2}\eta^{-1}\right.$$
$$-\frac{e^{2\iota\beta_e}}{2}\sum_{k=1}^{N}kb_k\eta^{-k-4}-\frac{1}{2}\sum_{k=1}^{N}kb_k\eta^{-k-2}-\frac{\bar{b}_1}{2}\sum_{k=1}^{N}kb_k\eta^{-k}-2e^{2\iota\beta_e}\bar{b}_2\sum_{k=1}^{N}kb_k\eta^{-k-1}$$
$$-e^{2\iota\beta_e}\sum_{k=1}^{N}\left(\sum_{p=1}^{N+k-3}\frac{p-k+3}{p-k+2}pb_p\bar{b}_{p-k+3}\right)\eta^{-k}-\sum_{k=1}^{N-1}\left(\sum_{p=k+1}^{N}\frac{2(p-k+1)^2}{(p-k+1)^2-1}pb_p\bar{b}_{p-k+1}\right)\eta^{-k}$$
$$\left.-e^{-2\iota\beta_e}\sum_{k=1}^{N-2}\left(\sum_{p=k+2}^{N}\frac{p-k-1}{p-k}pb_p\bar{b}_{p-k-1}\right)\eta^{-k}\right],\tag{A.11}$$

with

$$\lambda_k=\sum_{l=k}^{N+2}l\left(b_l\bar{\alpha}_{l-k+1}+\alpha_l\bar{b}_{l-k+1}\right),\qquad k=1,\dots,N+2.\tag{A.12}$$

The constant terms are arranged as

$$C=R\sum_{k=1}^{N}kb_k\bar{\alpha}_{k+1}+R\sum_{k=1}^{N-1}k\alpha_k\bar{b}_{k+1}-RA\bar{b}_1-R\sigma_{e0}\left[2e^{2\iota\beta_e}\bar{b}_2-\sum_{k=1}^{N-1}\frac{2(k+1)^2}{(k+1)^2-1}kb_k\bar{b}_{k+1}\right.$$
$$\left.-e^{-2\iota\beta_e}\sum_{k=1}^{N}\frac{k-1}{k}kb_k\bar{b}_{k-1}-e^{2\iota\beta_e}\sum_{k=1}^{N-3}\frac{k+3}{k+2}kb_k\bar{b}_{k+3}\right]-2\mu\bar{U}_0\equiv 0.\tag{A.13}$$

**Note**: Substituting $g_1=b_1+\sum_{l=1}^{N-2}lg_{l+2}\bar{b}_l$ from (50) into (53) yields

$$-2\mu\frac{\bar{U}_0}{R}=-\sum_{l=1}^{N-1}l\bar{g}_{l+1}\alpha_l-A\left(\bar{b}_1+\sum_{l=1}^{N-2}l\bar{g}_{l+2}\,b_l\right)-2b_2\sigma_{e0}e^{2\iota\beta_e},\tag{A.14}$$

combined with (51) and (52), $\alpha_{k+1}$ can be written as

$$\alpha_{k+1}=-\bar{A}g_{k+2}-\sigma_{e0}\left[\frac{2(k+1)^2}{(k+1)^2-1}b_{k+1}+\frac{k-1}{k}e^{2\iota\beta_e}b_{k-1}+\frac{k+3}{k+2}e^{-2\iota\beta_e}b_{k+3}\right]+\sum_{l=1}^{N-k-2}l\,\bar{\alpha}_lg_{k+l+2}.\tag{A.15}$$

Inserting (A.14) and (A.15) into (A.13), we obtain

$$C=R\sum_{k=1}^{N}kb_k\sum_{l=1}^{N-k-2}l\,\alpha_l\bar{g}_{k+l+2}+R\sum_{k=1}^{N-1}k\alpha_k\bar{b}_{k+1}-R\sum_{k=1}^{N-1}k\bar{g}_{k+1}\alpha_k$$
$$=R\sum_{k=1}^{N-1}k\left(b_k\sum_{l=1}^{N-k-2}l\alpha_l\bar{g}_{k+l+2}-\alpha_k\sum_{l=1}^{N-k-2}lb_l\bar{g}_{k+l+2}\right)\equiv 0,\tag{A.16}$$

where use is made of

$$g_{k+1}=b_{k+1}+\sum_{l=1}^{N-k-2}lg_{k+l+2}\bar{b}_l\tag{A.17}$$

from (50).